\documentclass[11pt]{article}
\usepackage{amssymb,latexsym,amsmath}
\usepackage[english]{babel}
\usepackage{xcolor}
\usepackage{bm}
\usepackage{amsmath,amscd}
\usepackage{graphicx}
\usepackage{newtxtext}
\usepackage{newtxmath}
\usepackage{geometry}
\usepackage{hyperref}
\hypersetup{
    colorlinks = true,
    urlcolor   = blue,
    citecolor  = black,
}

\newcommand{\RomanNumeralCaps}[1]
\linenumbers

\title{Variational theory of Cosserat arches and affine tensors}

\author{\textbf{G. de Saxc\'e} \\
Univ. Lille, CNRS, Centrale Lille, UMR 9013 – LaMcube \\
Laboratoire de m\'ecanique multiphysique multi\'echelle, \\
F-59000, Lille, France, Email: gery.de-saxce@univ-lille.fr}

\begin{document}
\maketitle

\begin{abstract}
Our purpose is to revisit the screw theory in light of the affine tensor formalism, introducing the co-momentum and momentum tensors. Our target-applications of the Euler-Poincar\'e equation are problems of mechanics such as the motion of the rigid body or the statics and the dynamics of Cosserat arches, in relation to the concept of momentum tensor. Using the framework of Ehresmann connections on the principal  bundle of affine frames, we show that the Euler-Poincar\'e equation means that the momentum tensor is parallel-transported.
\end{abstract}           

{\bf Keywords:} affine tensors, calculus of variations, Lie groups, screw theory, equations of motion, rods, beams, arches, shells. 

\vspace{0.2cm}

{\bf MSC Codes }  22E70; 74K25, 74K10; 
83C10







\section{Introduction}

The moment of a force, due to Archimedes, is a fundamental concept of the mechanical science. Its first mathematical formalization is due to Robert Stawell Ball \cite{Ball 1876} in the form of the theory of screws, algebraic objects composed of two dual vectors, declined in twists to describe the motion of a rigid body and wrenches to represent the action of a force acting on a rigid body. 
In the French literature \cite{Peres 1953}, the corresponding axiomatic is that of \textit{torseurs}. However, although the twist uses a translation vector, very few interests have been taken in wondering about the affine nature of this object. 

These notions of screws or \textit{torseurs} can be presented with a minimal background of vector calculus. At a higher mathematical level, another not less overlooked keystone of the Mechanics is the concept of continuous medium, especially organized around the tensor calculus which stems from Cauchy's works about the stresses (\cite{Cauchy 1823}, \cite{Cauchy 1827}). The general rules of this calculus were introduced by Ricci-Curbastro and Levi-Civita \cite{Ricci 1901}. They are concerned by the tensors that we shall call `linear tensors' insofar as their components are modified by means of linear base changes, then of regular linear transformations, elements of the linear group. 

According to \'{E}lie Cartan's ideas, a tensor can be also affine, projective, conformal, depending on the choice of the group. This is the point of view adopted in this work. The importance in Mechanics of the concept of affine tensor was originally pointed out and developed by Wlodzimierz Tulczyjew and his school \cite{Tulczyjew 1988, Grabowska 2004}. In \cite{Souriau 1997a}, Jean-Marie Souriau highlighted the affine character of many features of the Mechanics and proposed a general approach called "Affine Mechanics" and based on tensor-distributions. Our first purpose in this article is to revisit the screw theory in light of the affine tensor formalism. 

In the classical calculus of variation, the argument of the functional is a vector field, that leads to the Euler-Lagrange equation. In the famous note \cite{Poincare 1901} published by Henri Poincar\'e in 1901, he made an important breakthrough by considering functionals of which the argument lives in a non Abelian Lie group. The stationary points are characterized by the Euler-Poincar\'e equation. This approach is also known in the literature as Hamel's formalism \cite{Hamel 1904}, of which a modern exposition with various applications can be found in \cite{Shi 2017}. In \cite {Holm 1998}, the attention is payed to applications with a semi-direct product of a Lie group and a vector space. In \cite{Marle 2013}, Euler-Poincar\'e equation, reformulated in intrinsic form, is expressed in terms of the Legendre map and the momentum map in symplectic mechanics.

The second purpose of the paper is its applications to mechanical problems such as the motion of the rigid body, the statics and the dynamics of 1D Cosserat media \cite{Cosserat, Cosserat 1909}, a generic term that covers, in context, various engineering applications such as beams, rods, strings and arches. From now on, we will use the term arch to insist on the fact that in the most general case the structural element can be curved. In structural mechanics, our approach is similar to the so-called theory of geometrically-exact beams or rods \cite{Simo 1985, Vu-Quoc 1995, Boyer 2022}, used now in particular in bio-mimetic robotics for the numerical  simulation of the motion of eel-like robots \cite{Boyer 2009}. Our modeling of the arch by a moving frame is similar to the method used in \cite{Ivey 2003} to study the equivalence of curves.  

The last purpose of the paper is to interpret the Euler-Poincar\'e equation in terms of covariant derivatives, considering a connection on a principal bundle of affine frames \cite{Kobayashi}. Our approach is closed to the  symplectic approach developed in (\cite{AffineMechBook}, Sections 17.9 to 17.11, page 387-397; \cite{de Saxce GSI 2019}) and the mathematical framework proposed in \cite{Castrillon 2000} that we particularize to Euclid's group. 

\vspace{0.3cm}

The paper is structured as follows.

The Section \ref{Section - Affine tensors} is a very quick survey of the affine tensors, the transformation laws of their components and their applications in Mechanics. 

In Section \ref{Section - Co-momentum tensors}, we define the co-momentum tensor as a generalization of the twist in the screw theory. We assign components to it thanks to an affine frame. We observe that the component system of a co-momentum tensor is an element of the Lie algebra of the affine group and their transformation law is the adjoint representation. Introducing an Euclidean structure, we define the Euclidean co-momentum tensors, giving an interpretation of the twist in terms of differential geometry. 

In Section \ref{Section - Momentum tensors}, we define the momentum tensor as a generalization of the wrench in the screw theory. We define the Euclidean momentum tensors and put them in duality with the Euclidean co-momentum tensors. We observe that their component systems are the elements of the dual of the Lie algebra of the group of affine transformations that preserve the Euclidean structure. The term \textit{momentum} is used as name of this tensor to refer to the fact that it is the value of the momentum map in symplectic mechanics. 

The previous formalism is illustrated by applications to the dynamics of rigid bodies in Section \ref{Section - Dynamics of rigid bodies} and to the statics  of arches in Section \ref{Section - Static of arches}, using the left Maurer-Cartan 1-form for the material description and its right counterpart for the spatial description. 

In Section \ref{Section - Euler-Poincare equation}, we revisit the Euler-Poincar\'e equation. The approach is general but, to fix the ideas, we discuss the particular case of the dynamics of rigid bodies. By derivation of the Lagrangian, we obtain the constitutive relation between the co-momentum and the momentum. We apply these equations to the statics of arches. 

The purpose of Section \ref{Section - Generalization to continuous media of arbitrary dimension} is to generalize the previous formalism to continuous media of arbitrary dimension, equal for instance to $2$ for the dynamics of arches and to $3$ for the statics of shells.

In Section \ref{Section - Principal connection and covariant derivative}, we recall the concepts of an Ehresmann connection on a principal bundle and the covariant derivative on an associated principal bundle. We show that the left Maurer-Cartan 1-form defines a flat connection on the principal bundle of frames. 

Section \ref{Section - Interpretation of the equations in terms of covariant derivatives} is devoted to the interpretation of the Euler-Poincar\'e equation in terms of covariant derivatives: for the stationary points of the functional, the momentum tensor is parallel-transported. We show also that the flat connection is torsion free.

\section{Affine tensors}
\label{Section - Affine tensors}

Affine tensors are maps that are affine or linear with respect to their arguments and for which the affine group $\mathbb{GA} (d)$ or any of its Lie subgroups $G$ acts on their components by linear or affine representation (thus classical tensors are particular cases of affine  tensors). For the Mechanics, typical examples of such subgroups $G$ are the symmetry groups of the Physics such as Poincar\'e group, Galilei group and Bargmann group, or Euclid group to modelize  the motion of a rigid body. The aim of this section is to give the basic notions of the affine tensor calculus useful in the sequel. To know more about affine tensors and bundles, the reader is referred to the publications of Tulczyjew, Urbañski and Grabowski \cite{Tulczyjew 1988, Grabowska 2004} and to the author's one \cite{de Saxce 2003, de Saxce 2011, AffineMechBook, de Saxce 2019, de Saxce 2024}.

\begin{figure}[ht!]
\centering
\includegraphics[scale=.60]{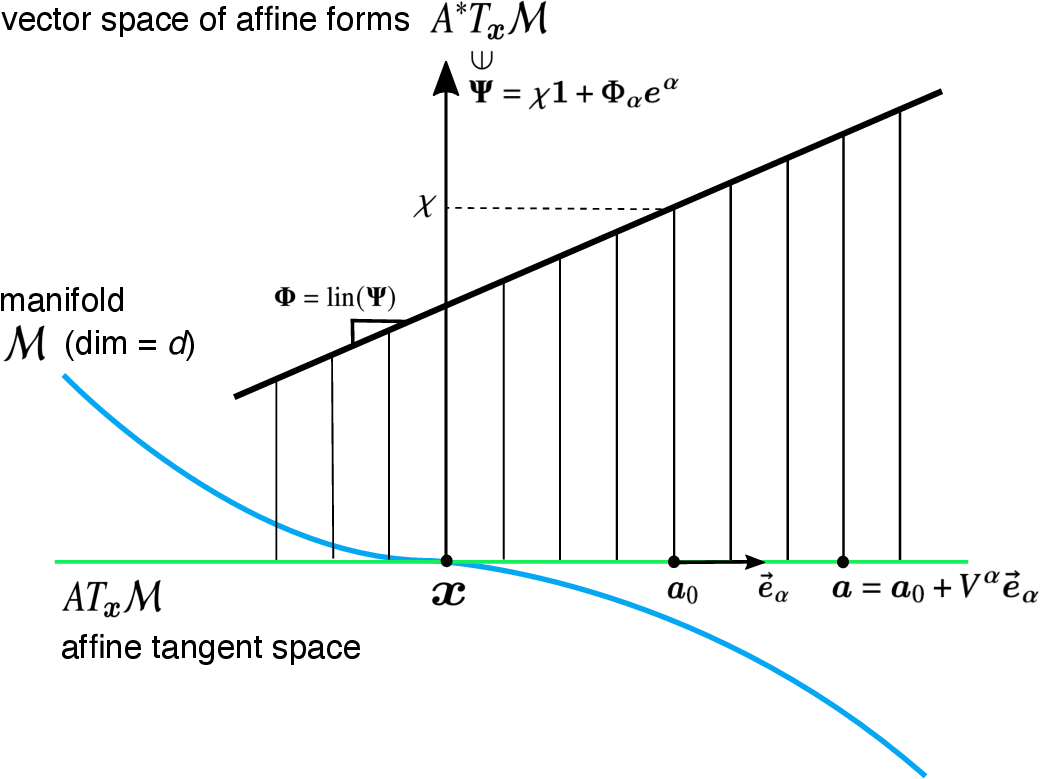}
\caption{Affine tensors: points and affine forms}
\label{fig points and affine forms}
\end{figure}

In a nutshell, we present the general setting. 
We consider a manifold $\mathcal{M}$ of dimension $d$ that may be the space ($ d = 3$)  or the space-time ($d = 4$). Our convention is that intrinsic, coordinate free objects are denoted by boldface letters while their components in local frames and charts are denoted by normal letters. 
$AT_{\bm{x}}\mathcal{M}$ is the affine space associated to the tangent vector space to $\mathcal{M}$ at $X$. $A^* T_{\bm{x}}\mathcal{M}$ is the vector space of the real-valued affine functions $\bm{\Psi}$ on $AT_{\bm{x}}\mathcal{M}$. They are called affine forms. The simplest affine tensors are (Figure \ref{fig points and affine forms}):
\begin{itemize}
    \item Firstly, the \textbf{points} $\bm{a}$  of the affine space. Similarly to vectors, they are 1-contravariant. By the choice of an affine frame, \textit{i.e.} an origin $\bm{a}_0$ and a basis $(\vec{\bm{e}}_\alpha)$ of $AT_{\bm{x}}\mathcal{M}$, decomposing the bound vector from the origin $\bm{a}_0$ to $\bm{a}$ in the basis
    \begin{equation}
       \bm{a} = \bm{a}_0 + V^\alpha \vec{\bm{e}}_\alpha
    \label{a = a_0 + V^alpha vec(e)_alpha}    
    \end{equation}
    we assign to $\bm{a}$ its components $V^\alpha$.
    \item Secondly, the \textbf{affine forms} $\bm{\Psi}$. Similarly to linear forms, they are 1-covariant. Decomposing it in the basis of $A^* T_{\bm{x}}\mathcal{M}$, \textit{i.e.} the constant function of value equal to 1, denoted $\bm{1}$, and the co-basis $(\bm{e}^\alpha)$ 
    \begin{equation}
        \bm{\Psi} = \chi \bm{1} + \Phi_\alpha \bm{e}^\alpha
    \label{Psi = chi 1 + Phi_alpha e^alpha}
    \end{equation}
    we assign to $\bm{\Psi}$ its components, the height $\chi$ at the origin and the components $\Phi_\alpha$ of the unique associated linear form $\bm{\Phi} = \mbox{lin} (\bm{\Psi})$, called linear  part of $\bm{\Phi}$.
\end{itemize}

Now, we present the affine tensors that will turn out to be the most relevant for the Mechanics and which have been previously studied by the author (for a survey, see \cite{AffineMechBook, de Saxce 2024} and, in French, \cite{de Saxce 2019})

\begin{itemize}
\item The \textbf{torsors} $\bm{\tau}$ that are 2-contravariant. A torsor is a bilinear and skew-symmetric function on the space of affine forms
$$\bm{\tau} (\bm{\Psi}, \hat{\bm{\Psi}} ) = - \bm{\tau} (\hat{\bm{\Psi}} ,\bm{\Psi})
$$
It is real or vector-valued. It is a mathematical object able to modelize the behaviour of material bodies. 
\item The \textbf{co-torsors} are 2-covariant. A co-torsor is a bi-affine and skew-symmetric function on the affine space $AT_{\bm{x}}\mathcal{M}$, real or vector-valued
$$\bm{\gamma} (\bm{a}, \hat{\bm{a}} ) = - \bm{\gamma} (\hat{\bm{a}} ,\bm{a})
$$
The co-torsors allow describing the kinematics and are in duality with the torsors (for more details, see \cite{AffineMechBook}, Section 5.1.3., page 76-80).
\end{itemize}


Now, we discuss the \textbf{transformation laws of affine tensors}, considering a change of affine frames  $(\bm{a}_0, (\vec{\bm{e}}_\alpha)) \longrightarrow (\bm{a}'_0, (\vec{\bm{e}}'_\beta)) $ given by the components $C'^\beta$ of the bound vector pointing from $\bm{a}'_0$ to $\bm{a}_0$  in the new basis, that can be stored in a column $C'$, and the transformation matrix $P$ of the basis change
\begin{eqnarray}
    \overrightarrow{\bm{a}'_0 \bm{a}_0} = C'^\beta \vec{\bm{e}}'_\beta, \qquad 
  \vec{\bm{e}}'_\beta = P^\alpha_\beta \vec{\bm{e}}_\alpha
\label{affine frame change}
\end{eqnarray}
Taking into account that $\mathcal{M}$ is of dimension $d$, we have
\begin{eqnarray}
     C' =\left( \begin{array} {c}
                       C'^1    \\
                       \vdots\\
                       C'^d   \\
                    \end {array} \right), \qquad 
   P =\left( {{\begin{array}{*{20}c}
                 P^1_1 \hfill & \ldots \hfill & P^1_d \hfill \\
                 \vdots \hfill & \ddots \hfill & \vdots \hfill \\
                 P^d_1 \hfill & \ldots \hfill & P^d_d \hfill \\
              \end{array} }} \right), \qquad
              C = -P \, C'
\label{C' = & P = & C =}
\end{eqnarray}
Thus —in matrix form— the laws of transformation of affine tensors are given 

\begin{itemize}
    \item for a point $\bm{a}$ of components $V^\alpha$ by
    $$V' = C' + P^{-1} V
    $$
    storing its components into the column $V$,
    \item for an affine form $\bm{\Psi}$ of components $(\chi, \Phi_\alpha)$ by
    $$\chi' = \chi + \Phi\,C,\qquad \Phi' = \Phi\,P 
    $$
    gathering the components $\Phi_\alpha$ into the row $\Phi$.
\end{itemize}
\vspace{0.3cm}
Now, combining the change of affine frames, we observe that the set of couples $(C, P)$ is the affine group
$$\mathbb{GA} (d) = \mathbb{R}^d \rtimes \mathbb{GL} (d)
$$
For a more rational organization of calculations, it is convenient to use a classical trick, the  linear representation on $\mathbb{R}^{d+1}$ of the affine group of $\mathbb{R}^d$
\begin{equation}
       \mathbb{GA} (d) \rightarrow \mathbb{GL} (d + 1): (C,P) \mapsto
        \tilde{P} =\left( {{\begin{array}{*{20}c}
                 1 \hfill & 0 \hfill \\
                 C \hfill & P \hfill \\
              \end{array} }} \right)   
\label{tilde(P) = ((1 0) (C P))}
\end{equation}
With this extra fifth dimension without physical meaning, we store the components of affine tensors into bigger column and row
\begin{equation}
     \tilde{V} =\left( \begin{array} {c}
                       1    \\
                       V    \\
                    \end {array} \right), \qquad 
 \tilde{\Psi} = (\chi \; \Phi)
\label{tilde(V) = (1 V) & tilde(Psi) = (chi Phi)}
\end{equation}
Then the transformation laws of the corresponding tensors take these simpler and compact forms
\begin{equation}
 \tilde{V} '= \tilde{P}^{-1} \tilde{V}, \qquad
      \tilde{\Psi}' = \tilde{\Psi}\,\tilde{P}
\label{tilde(Psi) = tilde(Psi)' P^(- 1)}    
\end{equation}

\section{Co-momentum tensors}
\label{Section - Co-momentum tensors}

\begin{figure}[ht!]
\centering
\includegraphics[scale=.60]{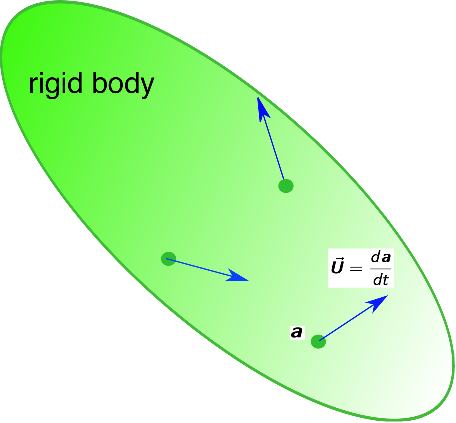}
\caption{Motion of a rigid body}
\label{fig Motion of a rigid body}
\end{figure}

We start with the simplest case where the co-momentum and momentum tensors are scalar-valued. One of our target application is the description of the motion of a rigid body. In the screw theory, a twist is an object composed of two dual vectors, a velocity and an angular velocity, that allows assigning to a point $\bm{a}$ of the body its velocity $\vec{\bm{U}} = d\bm{a} /dt$ (Figure \ref{fig Motion of a rigid body}). On this base, we consider an affine map $\bar{\bm{\theta}}$ from the tangent affine space $A T_{\bm{x}} \mathcal{M}$ into the tangent vector space $T_{\bm{x}} \mathcal{M}$ 
$$ \bar{\bm{\theta}}: A T_{\bm{x}} \mathcal{M}
\rightarrow T_{\bm{x}} \mathcal{M}: \bm{a} \mapsto 
\vec{\bm{U}} = \bar{\bm{\theta}} (\bm{a})
$$
It is a vector valued affine tensor which can be identified with the scalar valued mixed $1$-covariant and $1$-contravariant affine tensor $\bm{\omega}$ defined by
\begin{equation}
      \bm{\theta} (\bm{\Phi}, \bm{a}) 
= \bm{\Phi} (\bar{\bm{\theta}}  (\bm{a}))
\label{omega (Phi, a) = Phi ( bar(omega) (a))}
\end{equation}
To assign components $(\Upsilon^\beta, K^\beta_\alpha)$ to $\bm{\theta}$, we decompose its arguments in the affine frame of $A T_{\bm{x}}  \mathcal{M}$ and the basis of $A^* T_{\bm{x}} \mathcal{M}$, next we take into account that  $\bm{\theta}$ is linear with respect to $\bm{\Phi}$ and affine with respect to $\bm{a}$
\begin{equation}
       \bm{\theta} (\bm{\Phi}, \bm{a}) 
= \bm{\theta} (\Phi_\beta \bm{e}^\beta, \bm{a}_0 + V^\alpha \vec{\bm{e}}_\alpha) 
= \Phi_\beta (\Upsilon^\beta + K^\beta_\alpha V^\alpha)
\label{omega (Phi, a) =}
\end{equation}  
where its components are 
$$ \Upsilon^\beta = \bm{\theta} (\bm{e}^\beta, \bm{a}_0), \qquad
   K^\beta_\alpha = \bm{\theta} ( \bm{e}^\beta, \vec{\mathbf{e}}_\alpha)
$$
Moreover pay attention to the following subtleties. 
Because $V^\alpha = e^\alpha (\overrightarrow{\bm{a}_0 \bm{a}})$, by convention 
\begin{equation}
    V^\alpha = \bm{e}^\alpha (\bm{a})
\label{V^alpha = e^alpha (a)}
\end{equation}
and, identifying $\vec{\bm{e}}_\alpha$ with an element of the bidual, the value of $\bm{\Phi}$ for $\vec{\bm{e}}_\beta$  is the value of $\vec{\bm{e}}_\beta$ for $\bm{\Phi}$ 
$$ \Phi_\beta V^\alpha 
= \bm{\Phi} (\vec{\bm{e}}_\beta) \bm{e}^\alpha (\vec{\bm{V}})
 = \vec{\bm{e}}_\beta (\bm{\Phi}) \bm{e}^\alpha (\bm{a})
 = (\vec{\bm{e}}_\beta \otimes \bm{e}^\alpha) \, (\bm{\Phi}, \bm{a}) 
$$
Besides
$$ \Phi_\beta = \Phi_\beta . 1
 = \vec{\bm{e}}_\beta (\bm{\Phi}) \, \bm{1} (\bm{a}) = (\vec{\bm{e}}_\beta \otimes \bm{1}) \, (\bm{\Phi}, \bm{a})
$$
Equation (\ref{omega (Phi, a) =}) becomes
$$ \bm{\theta} (\bm{\Phi}, \bm{a}) 
= \Upsilon^\beta (\vec{\bm{e}}_\beta \otimes \bm{1}) \, (\bm{\Phi}, \bm{a})
 + K^\beta_\alpha (\vec{\bm{e}}_\beta \otimes \bm{e}^\alpha) \, (\bm{\Phi}, \bm{a})
$$
As $\bm{\Phi}$ and $\bm{a}$ are arbitrary, we obtain the decomposition in bases of $ T_{\bm{x}} \mathcal{M}$ and $A^* T_{\bm{x}} \mathcal{M}$
\begin{equation}
    \bm{\theta}  = \vec{\mathbf{e}}_\beta \otimes (\Upsilon^\beta \bm{1} + K^\beta_\alpha \bm{e}^\alpha ) 
\label{decomposition of the scalar-valued co-momentum}
\end{equation}
Introducing the $d$-column $\Upsilon$ collecting the $\Upsilon^\beta$ and the $d\times d$ matrix $K$ of elements $K^\beta_\alpha$, we obtain the matrix form of (\ref{omega (Phi, a) =})
\begin{equation}
     \bm{\theta} (\bm{\Phi}, \bm{a})   
= \Phi \, (\Upsilon + K \, V)
\label{theta (Phi, a) = Phi (Upsilon + K V)}
\end{equation}
that reads in compact form
$$  \bm{\theta} (\bm{\Phi}, \bm{a}) 
= \Phi\,\theta\,\tilde{V}
$$ 
by introducing the $d \times (d+1)$ matrix
\begin{equation}
   \theta = \left( \Upsilon \, K \right)
\label{tilde(omega) = (Upsilon K)} 
\end{equation}
According to the usual rules of the tensor calculus, the transformation laws  $\Phi' = \Phi \, P$ of the linear forms and (\ref{tilde(Psi) = tilde(Psi)' P^(- 1)}) of the points induce
\begin{equation}
     \theta
     = P\,\theta'\,\tilde{P}^{- 1}
\label{tilde(omega) = P tilde(omega') tilde(P)^(-1)}
\end{equation}
Owing to (\ref{tilde(P) = ((1 0) (C P))}), (\ref{tilde(V) = (1 V) & tilde(Psi) = (chi Phi)}) and (\ref{tilde(omega) = (Upsilon K)}), it is equivalent to
\begin{equation}
   \Upsilon = P \, (\Upsilon' - K' \, P^{-1} C),\qquad 
   K = P \, K' \, P^{-1} 
   \label{transformation law of co-momentum tensor} 
\end{equation}
Finally, comparing (\ref{omega (Phi, a) = Phi ( bar(omega) (a))}) and (\ref{omega (Phi, a) =}), we see that the affine map $\bar{\bm{\theta}}$ from $A T_{\bm{x}} \mathcal{M}$ into  $T_{\bm{x}} \mathcal{M}$ is represented in the affine frame and the basis by
\begin{equation}
     \bar{\theta} (\tilde{V}) 
= \Upsilon + K \, V = \theta \, \tilde{V}
\label{bar(omega) (a) = Upsilon + K V}
\end{equation}

\subsection{Adjoint representation}

On the other hand, have a look to the Lie algebra $\mathfrak{ga} (d)$ of the affine group $\mathbb{GA} (d)$, that is the set of infinitesimal generators $Z = d\mathsf{g} = (dC, dP)$. 
We know that this group acts on its Lie algebra by the \textbf{adjoint representation}
$$ Ad (\mathsf{g}): \mathfrak{ga} (d) \rightarrow \mathfrak{ga} (d): Z' \mapsto Z = Ad (\mathsf{g})\,Z' = \mathsf{g}\,Z'\,\mathsf{g}^{- 1}
$$ 
Using the  linear representation of the affine group of Section \ref{Section - Affine tensors}, any infinitesimal generator $Z$ is represented by 
$$ \tilde{Z} = d\tilde{P} 
             = d \left( {{\begin{array}{*{20}c}
                     1 \hfill & 0 \hfill \\
                     C \hfill & P \hfill \\
   \end{array} }} \right)
             =   \left( {{\begin{array}{*{20}c}
                     0  \hfill & 0  \hfill \\
                     dC \hfill & dP \hfill \\
   \end{array} }} \right)\ .
$$
Then $\tilde{Z} = \tilde{P}\,\tilde{Z}'\,\tilde{P}^{- 1}$ leads to
\begin{equation}
   dC = P\,(dC' - dP'\,P^{- 1} C),\qquad dP = P\,dP'\,P^{- 1}\ .
\label{dC = P (dC' - dP' P^(-1) C) & dP = P dP' P^(-1)} 
\end{equation}
Comparing to (\ref{transformation law of co-momentum tensor}), we see that \textbf{the transformation law of co-momentum tensors is the adjoint representation}
\begin{equation}
     \theta = Ad (\mathsf{g})\,\theta'
\label{theta = Ad (g) theta'}
\end{equation}
In other words, \textbf{the couple} (\ref{tilde(omega) = (Upsilon K)})  \textbf{collecting the components of the co-momentum} $\bm{\theta}$ \textbf{is an element of the Lie algebra} $\mathfrak{ga} (d)$. Note also that the set of co-momentum tensors is a vector space of dimension $d \times (d+1)$ and that the adjoint representation of the affine group induces by restriction the corresponding representation on every Lie subgroup.

\subsection{Euclidean co-momentum tensors}

In the screw theory, a twist represents the motion of a rigid body, that is a body of which the distance between its points is invariant. To define it in the present formalism, we need to introduce a metric tensor that leads to the concept of Euclidean co-momentum. We consider now that the manifold $\mathcal{M}$ is Riemannian, and thus $T_{\bm{x}} \mathcal{M}$ is endowed with an Euclidean structure defined by a covariant metric tensor. Denoting $\left\langle  \bullet, \bullet \right\rangle$ the scalar product, the \textbf{adjoint} $\bm{A}^\star$ of an endomorphism $\bm{A}$ of $T_{\bm{x}} \mathcal{M}$ is such as 
$$ \forall \,  \vec{\bm{U}} , \vec{\bm{V}} \in T_{\bm{x}} \mathcal{M}, \qquad 
\left\langle \bm{A}^\star \vec{\bm{U}} , \vec{\bm{V}} \right\rangle = \left\langle  \vec{\bm{U}} , \bm{A} \, \vec{\bm{V}} \right\rangle
$$
$\bm{A}$ is skew-adjoint if $\bm{A}^\star = - \bm{A}$. As for the affine forms, we can define the linear part of the co-momentum (see \cite{AffineMechBook}, Section 14.2.2.)
\begin{equation}
     (\mbox{lin} (\bm{\theta})) (\bm{\Phi}, \overrightarrow{\bm{a}_0 \bm{a}}) = \bm{\theta} (\bm{\Phi}, \bm{a}) - \bm{\theta} (\bm{\Phi}, \bm{a}_0) 
\label{lin (theta) =}
\end{equation}
which is linear with respect to its second argument because the co-momentum is affine with respect to its last argument. 
Thus there is an endomorphism of $T_{\bm{x}} \mathcal{M}$ such that
$$ (\mbox{lin} (\bm{\theta})) (\bm{\Phi}, \vec{\bm{V}}) 
= \bm{\Phi} (\bm{K} (\vec{\bm{V}}))
$$
Taking into account (\ref{theta (Phi, a) = Phi (Upsilon + K V)}) and (\ref{lin (theta) =}), we see that it is represented in a basis by the matrix $K$
$$ (\mbox{lin} (\bm{\theta})) (\bm{\Phi}, \vec{\bm{V}}) 
= \Phi \, (\Upsilon + K \, V) - \Phi \, \Upsilon 
= \Phi \,  K \, V
$$
To a Riemannian manifold $\mathcal{M}$ we can associate the Lie subgroup $G$ of affine transformations $\mathsf{g} = (C, P)$ which leave invariant this Euclidean structure, in other words Euclid's group in dimension $d$  
$$\mathbb{SE} (d) 
= \mathbb{R}^d \rtimes \mathbb{SO} (d)
$$
The infinitesimal generators $Z = (dC, dP)$ of its Lie algebra $\mathfrak{se} (d)$ are such that $dP$ is skew-adjoint. This motivates the following definition: we say that \textbf{the co-momentum} $\bm{\theta}$ \textbf{is Euclidean} if $\bm{K}$ is skew-adjoint, and $K$ as well.

\vspace{0.3cm}

On the other hand, for a given origin $\bm{a}_0$, $\bm{\theta} (\bm{\Phi}, \bm{a}_0)$ is a linear function of the first argument, then there exists a vector $\vec{\bm{\Upsilon}}_{\bm{a}_0}$ of $T_{\bm{x}} \mathcal{M}$ such that
$$ \bm{\theta} (\bm{\Phi}, \bm{a}_0) = \bm{\Phi} \, \vec{\bm{\Upsilon}}_{\bm{a}_0} 
$$
Owing to (\ref{theta (Phi, a) = Phi (Upsilon + K V)}), the vector $\vec{\bm{\Upsilon}}_{\bm{a}_0} $ is represented in the affine frame of origin $\bm{a}_0$ by the $d$-column $\Upsilon$. 

\vspace{0.3cm}

In a nutshell, $\bm{\theta}$ is characterized by $\vec{\bm{\Upsilon}}_{\bm{a}_0}$ and $\bm{K}$. 
The couple (\ref{tilde(omega) = (Upsilon K)})  collecting the components of the co-momentum $\bm{\theta}$ is an element of the Lie algebra $\mathfrak{se} (d)$ of $\mathbb{SE} (d) $. The set of Euclidean co-momenta is a vector space of dimension $d + d \, (d - 1) / 2$.

\section{Momentum tensors}
\label{Section - Momentum tensors}

We recall some basic notions introduced in (\cite{AffineMechBook}, Section 16.3).
Let us consider a linear map $\bar{\bm{\mu}}$ from the space $T^*_{\bm{X}} \mathcal{M}$ of linear forms into the one $A^* T_{\bm{x}} \mathcal{M}$ of affine forms. It is a vector valued affine tensor which can be identified with the scalar valued mixed $1$-covariant and $1$-contravariant affine tensor $\bm{\mu}$ defined by
$$   \bm{\mu} (\vec{\mathbf{V}}, \bm{\Psi}) = (\bar{\bm{\mu}} (\mathbf{\Psi})) \vec{\mathbf{V}}\ .
$$
To assign components $(\Pi_\beta, L^\alpha_\beta )$ to $\bm{\mu}$, we decompose its arguments  in the bases of $T_{\bm{x}}  \mathcal{M}$ and $A^* T_{\bm{x}} \mathcal{M}$, next we take into account the bilinearity
\begin{equation}
     \bm{\mu} (\vec{\mathbf{V}}, \bm{\Psi}) 
               = \bm{\mu} (V^\beta \vec{\mathbf{e}}_\beta, \chi \bm{1} + \Phi_\alpha \bm{e}^\alpha) 
               =  (\chi\,\Pi_\beta + \Phi_\alpha\,L^\alpha_\beta)\,V^\beta
\label{mu(vec(V),Psi) =}
\end{equation}
where
$$ \Pi_\beta = \bm{\mu} (\vec{\mathbf{e}}_\beta, \bm{1}),\qquad   
   L^\alpha_\beta = \bm{\mu} (\vec{\mathbf{e}}_\beta, \bm{e}^\alpha) \ .
$$ 
Moreover pay attention to the following subtleties. $\Phi_\alpha$
               is the value of $\bm{\Phi}$  for the basis vector $\vec{\bm{e}}_\alpha$ but, as $\bm{\Phi}$ is the linear part of $\bm{\Psi}$, by convention it is the value of $\bm{\Psi}$ for $\vec{\bm{e}}_\alpha$ 
               \begin{equation}
                  \bm{\Phi}  (\vec{\bm{e}}_\alpha) =     \bm{\Psi} (\vec{\bm{e}}_\alpha)
               \label{Phi (vec(e)_alpha) = Psi  (vec(e)_alpha)}
               \end{equation}
               and, identifying $\vec{\bm{e}}_\alpha$ with an element of the bidual, it is the value of $\vec{\bm{e}}_\alpha$ for $\bm{\Psi}$ 
$$\Phi_\alpha V^\beta 
= \bm{\Phi}  (\vec{\bm{e}}_\alpha) \, \bm{e}^\beta (\vec{\mathbf{V}})
    = \bm{\Psi} (\vec{\bm{e}}_\alpha) \, \bm{e}^\beta (\vec{\mathbf{V}})
 =  \vec{\bm{e}}_\alpha (\bm{\Psi})\, \bm{e}^\beta (\vec{\mathbf{V}}) 
 = (\bm{e}^\beta \otimes \vec{\bm{e}}_\alpha  ) (\vec{\mathbf{V}}, \bm{\Psi})  
 $$
Similarly, identifying the origin $\bm{a}_0$ with an element of the bidual, $\chi$ is the value of $\bm{a}_0$ for $\bm{\Psi}$
$$\chi =\bm{\Psi}  (\bm{a}_0) = \bm{a}_0 (\bm{\Psi}) 
$$
Equation (\ref{mu(vec(V),Psi) =}) becomes
$$ \bm{\mu} (\vec{\mathbf{V}}, \bm{\Psi}) 
=  \Pi_\beta (\bm{e}^\beta \otimes \bm{a}_0) (\vec{\mathbf{V}}, \bm{\Psi}) 
+ L^\alpha_\beta\,  (\bm{e}^\beta \otimes \vec{\bm{e}}_\alpha) (\vec{\mathbf{V}}, \bm{\Psi}) 
$$
As $\vec{\mathbf{V}}$ and $\bm{\Psi}$ are arbitrary,
we obtain the decomposition 
\begin{equation}
     \bm{\mu} = \bm{e}^\beta \otimes\,(\Pi_\beta \, \bm{a}_0 
              + L^\alpha_\beta\,  \vec{\bm{e}}_\alpha)
\label{decomposition of the scalar-valued momentum}
\end{equation}
Introducing the $d$-row $\Pi$ collecting the $\Pi_\beta$ and the $d\times d$ matrix $L$ of elements $L^\alpha_\beta$, we obtain the matrix form of (\ref{mu(vec(V),Psi) =})
\begin{equation}
     \bm{\mu} (\vec{\mathbf{V}}, \bm{\Psi})  = (\chi\,\Pi + \Phi\,L)\,V 
\label{mu (vec(V) Psi) = (chi Pi + Phi L) V}
\end{equation}
that reads in compact form
$$ \bm{\mu} (\vec{\mathbf{V}}, \bm{\Psi})  = \tilde{\Psi}\,\mu\,V\,
$$ 
by introducing the $(d+1) \times d$ matrix
\begin{equation}
   \mu = \left( {{\begin{array}{c}
                             \Pi \\
                             L 
                          \end{array} }} \right)\ .
\label{tilde(mu) = col(F, L)} 
\end{equation}

\subsection{Euclidean momentum tensors}

The affine forms $\bm{\Psi}$ which vanish at a given origin $\bm{a}_0$ are such that
$$ \bm{\Psi} (\bm{a}) = (\mbox{lin} (\bm{\Psi})) (\overrightarrow{\bm{a}_0 \bm{a}})  = \bm{\Phi} (\overrightarrow{\bm{a}_0 \bm{a}}) 
$$
Their set $A^*_{\bm{a}_0} T_{X} \mathcal{M}$ is a subspace of $A^* T_{X} \mathcal{M}$ of dimension $d$. Introducing the map
$$ p_{\bm{a}_0}: AT_{X} \mathcal{M} \rightarrow T_{X} \mathcal{M}: \bm{a} \mapsto \overrightarrow{\bm{a}_0 \bm{a}}
$$
the pullback 
$$ (p_{\bm{a}_0})^*: 
T^*_{X} \mathcal{M} \rightarrow A^*_{\bm{a}_0} T_{X} \mathcal{M} : \bm{\Phi} \mapsto
 \bm{\Psi} = \bm{\Phi} \circ p_{\bm{a}_0} 
$$
is an isomorphism. The map $\bm{\mu}_{\bm{a}_0}$ defined by
$$ \bm{\mu}_{\bm{a}_0} (\vec{\bm{V}}, \bm{\Phi}) 
   = \bm{\mu} (\vec{\bm{V}}, (p_{\bm{a}_0})^* \bm{\Phi})  
$$
is bilinear from $T_{X} \mathcal{M} \times T^*_{X} \mathcal{M}$ into $\mathbb{R}$, then there exists an endomorphism 
$$ \bm{\Lambda}_{\bm{a}_0}: T^*_{X} \mathcal{M} \rightarrow
T^*_{X} \mathcal{M}: \bm{\Phi} \mapsto \tilde{\bm{\Phi}} 
= \bm{\Lambda}_{\bm{a}_0} (\bm{\Phi})
$$
such that
\begin{equation}
     \bm{\mu}_{\bm{a}_0} (\vec{\bm{V}}, \bm{\Phi}) 
   =  (\bm{\Lambda}_{\bm{a}_0} (\bm{\Phi}) ) 
   \vec{\bm{V}}
\label{mu_a0 (vec(V), Phi) =(Lambda_a0 (Phi)) vec(V)}
\end{equation}
and represented in the dual basis $(\bm{e}^\alpha)$ by the $d \times d$ matrix $L$
$$ \tilde{\Phi} =  \Phi \, L
$$
Its transpose is the endomorphism 
$$ \bm{L}_{\bm{a}_0}: T_{X} \mathcal{M} \rightarrow
T_{X} \mathcal{M}: \vec{\bm{V}} \mapsto \vec{\tilde{\bm{V}}}
= \bm{L}_{\bm{a}_0} (\vec{\bm{V}})
$$
represented in the basis $(\vec{\bm{e}}_\alpha)$ by the same matrix $L$
$$ \tilde{V} = L \, V
$$
because vectors are represented by $d$-columns and linear forms by $d$-rows. In fact, $L$ is the matrix that gathers  components $L^\alpha_\beta$ since using (\ref{mu (vec(V) Psi) = (chi Pi + Phi L) V}) with $\chi = \bm{\Psi} (\bm{a}_0) = 0$ gives
$$ \bm{\mu}_{\bm{a}_0} (\vec{\mathbf{V}}, \bm{\Phi})  = \Phi\,L\,V 
$$
which represents (\ref{mu_a0 (vec(V), Phi) =(Lambda_a0 (Phi)) vec(V)}). We say that \textbf{the momentum} $\bm{\mu}$ \textbf{is Euclidean} if $\bm{L}_{\bm{a}_0}$ is skew-adjoint, and $L$ as well.

\vspace{0.3cm}

On the other hand, the set $A^*_c T_{X} \mathcal{M}$ of constant affine forms is a subspace of $A^* T_{X} \mathcal{M}$ of dimension $1$, isomorphic to $\mathbb{R}$ by $\chi  \mapsto \bm{\Psi} = \chi \, \bm{1}$. Let us consider
$$ \bm{\mu}_c (\vec{\bm{V}}, \chi)  = \bm{\mu} (\vec{\bm{V}}, \chi \, \bm{1}) = \chi \, \bm{\mu} (\vec{\bm{V}}, \bm{1})
$$
where the scalar $\bm{\mu} (\vec{\bm{V}}, \bm{1})$ is linear with respect to $\vec{\bm{V}}$. There is a linear form $\bm{\Pi} \in T^*_{X} \mathcal{M}$ such that 
$$ \bm{\mu} (\vec{\bm{V}}, \bm{1}) = \bm{\Pi}  \, \vec{\bm{V}}
$$
Owing to (\ref{mu (vec(V) Psi) = (chi Pi + Phi L) V}), the linear form $\bm{\Pi}$ is represented in the dual basis $(\bm{e}^\alpha)$ by the $d$-row $\Pi$. 

\vspace{0.3cm}

In summary, based on the direct sum
$$ A^* T_{X} \mathcal{M} = 
A^*_c T_{X} \mathcal{M}
\oplus 
A^*_{\bm{a}_0} T_{X} \mathcal{M}
$$
$\bm{\mu}$ is characterized by $\bm{\Pi}$ and $\bm{L}_{\bm{a}_0}$. The set of Euclidean momenta is a vector space of same dimension $d + d \, (d - 1) / 2$ as the one of the space of co-momenta.

\subsection{Coadjoint representation}

Using the characterization of the co-momentum 
$\bm{\theta}$ by the couple $(\vec{\bm{\Upsilon}}_{\bm{a}_0}, \bm{K})$ and of the momentum $\bm{\mu}$ by the couple $(\bm{\Pi}, \bm{L}_{\bm{a}_0})$, we define the \textbf{dual pairing of Euclidean co-momenta and momenta} by the expression 
$$ \bm{\mu} \, \bm{\theta} = \bm{\Pi} \, \vec{\bm{\Upsilon}}_{\bm{a}_0} - \frac{1}{2} Tr (\bm{L}_{\bm{a}_0}  \bm{K})
$$
in which the last term is meaningful because both $\bm{L}_{\bm{a}_0}$ and $\bm{K}$ are endomorphisms of $T_{\bm{x}} \mathcal{M}$. It is represented in an affine frame of $AT_{\bm{x}} \mathcal{M}$ and a basis of $A^*T_{\bm{x}} \mathcal{M}$ by 
\begin{equation}
     \bm{\mu} \, \bm{\theta} 
     = \mu \, \theta
=  \Pi \, \Upsilon
 - \frac{1}{2} \, Tr (L \, K)
\label{mu omega = Tr (tilde(mu) tilde(omega)) = Upsilon Pi + Tr (L K)}
\end{equation}
The adjoint representation induces the \textbf{coadjoint representation} of $\mathbb{SE} (d)$ in $(\mathfrak{se} (d))^*$ defined by
$$ (Ad^* (\mathsf{g})\,\mu')\,Z = \mu'\,(Ad (\mathsf{g}^{- 1})\,Z)\ .
$$
Owing to (\ref{mu omega = Tr (tilde(mu) tilde(omega)) = Upsilon Pi + Tr (L K)}), one finds that the coadjoint representation 
$$ Ad^* (\mathsf{g}) : (\mathfrak{se} (d))^* \rightarrow (\mathfrak{se} (d))^* : \mu' \mapsto \mu = Ad^* (\mathsf{g})\,\mu' 
$$
is given by
\begin{equation}
       \Pi = \Pi'\,P^{-1},\qquad 
       L = P\, L' \, P^{-1}
          + C\, \Pi'\, P^{-1} - (\Pi'\, P^{-1})^* C^*
\label{coadjoint representation}
\end{equation}
Then \textbf{the coadjoint action gives the transformation law of Euclidean momentum tensors}
\begin{equation}
    \mu = Ad^* (\mathsf{g})\,\mu'
\label{mu = Ad^* (g) mu'}
\end{equation}
In other words, \textbf{the couple} (\ref{tilde(omega) = (Upsilon K)})  \textbf{collecting the components of the Euclidean momentum} $\bm{\mu}$ \textbf{is an element of the dual} $(\mathfrak{se} (d))^*$ \textbf{of the Lie algebra}.

\section{Affine frames and co-frames}
\label{SubSection - Affine frames and co-frames}

An affine frame $f$ can be seen as a linear map of argument in $\mathbb{R}^{d + 1}$ and value in $AT_{\bm{x}} \mathcal{M}$ such that the decomposition (\ref{a = a_0 + V^alpha vec(e)_alpha}) of a point $\bm{a}$ is given by
$$ \bm{a} = f \, \tilde{V} 
  = ( \bm{a}_0 ,\vec{\bm{e}}_1,\ldots,\vec{\bm{e}}_d )\,
  \left( \begin{array} {c}
                       1      \\
                       V^1    \\
                       \vdots\\
                       V^d   \\
                    \end {array} \right)  
$$
Similarly, an \textbf{affine co-frame} $f^*$, \textit{i.e.} a basis of $A^* T_{\bm{x}} \mathcal{M}$, can be seen as an isomorphism from $AT_{\bm{x}} \mathcal{M}$ into $\mathbb{R}^{d + 1}$ such that the decomposition (\ref{Psi = chi 1 + Phi_alpha e^alpha}) of an affine form $\bm{\Psi}$ is given by
$$ \bm{\Psi} =  \tilde{\Psi} \, f^*
  = ( \chi ,\Phi_1,\ldots,\Phi_d )\,
  \left( \begin{array} {c}
                       \bm{1}      \\
                       \bm{e}^1    \\
                       \vdots\\
                       \bm{e}^d   \\
                    \end {array} \right)  
$$
As 
$\bm{\Psi} (\bm{a}) = \tilde{\Psi} \, \tilde{V}$,
the map $f^*$ is a retraction of $f$
$$ f^* f = 1_{\mathbb{R}^{d + 1}}
$$
that itemizes into
$$ \bm{1} (\bm{a}_0) = 1, \qquad 
   \bm{1} (\vec{\bm{e}}_\beta) = 0, \qquad
   \bm{e}^\alpha (\bm{a}_0) = 0, \qquad
   \bm{e}^\alpha (\vec{\bm{e}}_\beta) = \delta^\alpha_\beta
$$
with the following interpretation: the first condition means that the value of the function $\bm{1}$ is $1$; the second is the result of the convention (\ref{Phi (vec(e)_alpha) = Psi  (vec(e)_alpha)}) applied to $\bm{\Psi} = \bm{1}$;  the third is a consequence of the convention (\ref{V^alpha = e^alpha (a)}) applied to $\bm{a} = \bm{a}_0$; the latter is a classical relationship of linear algebra. By these conditions, the frame $f$ and the co-frame $f^*$ can be deduced one from each other.

The affine frames (resp. co-frames) of which the basis (resp. co-basis) is orthonormal are called \textbf{Euclidean frames} (resp. \textbf{Euclidean co-frames}). In a change of Euclidean frames (resp. co-frame, the components of Euclidean co-momenta (resp. momenta) are changed according to (\ref{theta = Ad (g) theta'}) (resp. (\ref{mu = Ad^* (g) mu'})) where $\mathsf{g} \in \mathbb{SE} (3)$. 

The set $\bm{x} \in \mathcal{M}$ of affine frames of $AT_{\bm{x}} \mathcal{M}$ for all is a $\mathbb{GA} (3)$-principal bundle $AT \mathcal{M}$ of base $\mathcal{M}$. The subbundle of Euclidean affine frames is a $\mathbb{SE} (3)$-principal bundle.

\section{Dynamics of rigid bodies}
\label{Section - Dynamics of rigid bodies}

\begin{figure}[ht!]
\centering
\includegraphics[scale=.60]{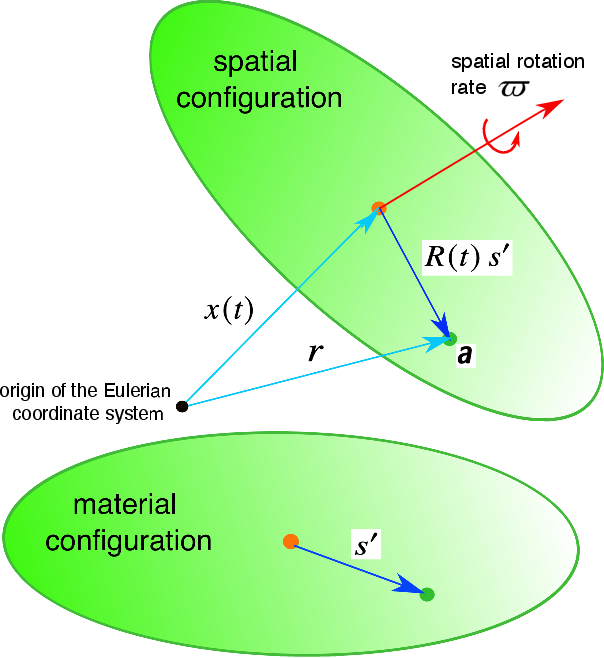}
\caption{Rigid body}
\label{fig Rigid body}
\end{figure}

Let $\mathcal{B}'_3 \subset \mathbb{R}^3$ be a rigid body and $x_0  \in \mathbb{R}^3$ be the position of its center of mass (Figure \ref{fig Rigid body}). In a reference frame, the position $r \in \mathbb{R}^3$ at time $t$ of a material point $s'\in \mathcal{B}'_3$ is given by
\begin{equation}
     r = R (t) \, s' + x (t)
\label{x = R (t) s' + x_0}
\end{equation}
Its Lagrangian (resp. Eulerian) coordinates are $s'^i$ (resp. $r^i$) and its velocity in the reference frame is
\begin{equation}
   \dot{r} = \dot{x}  + \varpi  \times (r - x)
\label{v = u} 
\end{equation}  
where $ \varpi$ is Poisson's vector
\begin{equation}
   \dot{R} = j (\varpi)\,R
\label{dot(R) Eulerian} 
\end{equation} 
$j (\varpi)$ being the skew-symmetric matrix such that $j (\varpi) v = \varpi \times v$. 
The overall motion of the body is defined by a path in Euclid's group 
$$ t \mapsto \mathsf{g} (t) = (x (t), R (t)) \in \mathbb{SE} (3) 
$$
We are going now to show by two different approaches how to construct the co-momentum and momentum tensors of the rigid body.

\subsection{Spatial description}

In this section and the following ones, the manifold $\mathcal{M}$ is the physical space of dimension 3.
Using the  linear representation of Euclid's group $\mathbb{SE} (3)$, Equation (\ref{x = R (t) s' + x_0}) reads
$$ \tilde{V} = \tilde{\mathsf{g}} \, \tilde{V}'
$$
with
\begin{equation}
      \tilde{V} =\left( \begin{array} {c}
                       1    \\
                       r    \\
                    \end {array} \right), \qquad 
    \tilde{\mathsf{g}} =\left( {{\begin{array}{*{20}c}
                 1 \hfill & 0 \hfill \\
                 x \hfill & R \hfill \\
              \end{array} }} \right), \qquad  
  \tilde{V}' =\left( \begin{array} {c}
                       1    \\
                       s'    \\
                    \end {array} \right)
\label{rigid body - tilde(V) = & tilde(g) = & tilde(V)' =}
\end{equation}
We know that the right Maurer-Cartan 1-form of $\mathbb{SE} (3)$ is a field of  1-forms $\vartheta_R$ on $\mathbb{SE} (3)$ valued in its Lie algebra $\mathfrak{se} (3)$, invariant by right translations and defined by 
\begin{equation}
     \theta = \vartheta_R (d \mathsf{g}) =
d \mathsf{g} \, \mathsf{g}^{-1}
\label{theta = vartheta_R (dg) = dg g^(-1)}
\end{equation}
We claim that \textbf{the components of the Euclidean co-momentum} $\bm{\theta}$ \textbf{of the rigid body are given by the value of} $\vartheta_R$ \textbf{for} $\dot{\mathsf{g}}$. 
Using the linear representation of the affine group, decomposing by blocks and owing to (\ref{dot(R) Eulerian}), the co-momentum $\bm{\theta}$ is represented by
$$ \tilde{\theta} = \dot{\tilde{\mathsf{g}}} \, \tilde{\mathsf{g}}^{-1} 
= \left( {{\begin{array}{*{20}c}
                 0 \hfill & 0 \hfill \\
                 \dot{x} \hfill & \dot{R} \hfill \\
              \end{array} }} \right) \,
  \left( {{\begin{array}{*{20}c}
                 1 \hfill & 0 \hfill \\
                 - R^T x \hfill & R^T \hfill \\
              \end{array} }} \right) 
 = \left( {{\begin{array}{*{20}c}
                 0 \hfill & 0 \hfill \\
                 \dot{x} - \varpi \times x \hfill &  j (\varpi ) \hfill \\
              \end{array} }} \right)
$$
and thus
$$ \vartheta_R (\dot{\mathsf{g}}) =
\left( \dot{x} - \varpi \times x, \; \; j (\varpi ) \right)
$$
By identification with  (\ref{tilde(omega) = (Upsilon K)}), we obtain the components of the co-momentum $\bm{\theta}$
\begin{equation}
     \Upsilon = \dot{x} - \varpi \times x, \qquad
   K = j (\varpi)
\label{rigid body - Eulerian - Upsilon = & K =}
\end{equation}
Owing to (\ref{v = u}), the component $\Upsilon$ is the velocity of the origin of the Eulerian coordinate system for which $r = 0$, arbitrarily chosen independently of the body. By the way, it is worth to observe that calculating (\ref{bar(omega) (a) = Upsilon + K V}) and comparing to (\ref{v = u}) yields
$$ \dot{r} = \bar{\theta} (r) 
        = \Upsilon + K \, r
$$
which gives the meaning of the map $\bar{\bm{\theta}}$ from $A T_{\bm{x}} \mathcal{M}$ into  $T_{\bm{x}} \mathcal{M}$ as assigning to a point $\bm{a}$ of the affine space its velocity 
$$ \vec{\bm{U}} = \frac{d \bm{a}}{dt}
$$

On the other hand, as the matrix $\mu$ collecting the components of the momentum tensor is an element of the dual $(\mathfrak{se} (3))^*$ of the Lie algebra of Euclid's group, $L$ is skew-symmetric and
we claim that
$$ \Pi = p^T, \qquad L =  j (l)
$$
where $p \in \mathbb{R}^3$ is the linear momentum and $l \in \mathbb{R}^3$ is the angular momentum. Owing to (\ref{mu omega = Tr (tilde(mu) tilde(omega)) = Upsilon Pi + Tr (L K)}) and (\ref{rigid body - Eulerian - Upsilon = & K =}), the value of the dual pairing of the momentum and the co-momentum is
\begin{equation}
     \bm{\mu} \, \bm{\theta}  
    = p \cdot \dot{x} + (l - x \times p ) \cdot \varpi
\label{mu omega = p cdot dot(x)_0 + (l - x_0 times p)}
\end{equation}
which, taking into account the decomposition of the angular momentum into orbital angular momentum and proper angular momentum (or spin) $l_{pr}$
$$ l =  x \times p + l_{pr}
$$
gives
$$ \bm{\mu} \, \bm{\theta}  
    = p \cdot \dot{x} + l_{pr} \cdot \varpi
$$
interpreted as the total power developed by the rigid body.

\subsection{Material description}

As for the Lagrange coordinates $s'^i$, all the quantities related to the material description are denoted with a prime. We know that the left Maurer-Cartan 1-form of $\mathbb{SE} (3)$ is a field of 1-forms $\vartheta_L$ on $\mathbb{SE} (3)$ valued in $\mathfrak{se} (3)$, invariant by left translations and defined by
$$ 
\vartheta_L (d \mathsf{g}) =
\mathsf{g}^{-1} d \mathsf{g} 
$$
Owing to the transformation law (\ref{theta = Ad (g) theta'}) of co-momenta , we see that the new components are the value of $\vartheta_L$ for $\dot{\mathsf{g}}$ 
\begin{equation}
     \theta' = Ad(\mathsf{g}^{-1}) \, \theta 
   = Ad(\mathsf{g}^{-1}) \, \vartheta_R (\dot{\mathsf{g}})
  = \vartheta_L (\dot{\mathsf{g}})
\label{theta' = Ad(g^(-1)) theta = Ad(g^(-1)) vartheta_R(dot(g)) = vartheta_L (dot(g))}
\end{equation}
They are represented by
$$ \tilde{\theta}' 
= \tilde{\mathsf{g}}^{-1} \dot{\tilde{\mathsf{g}}} 
=   \left( {{\begin{array}{*{20}c}
                 1 \hfill & 0 \hfill \\
                 - R^T x \hfill & R^T \hfill \\
              \end{array} }} \right) \,
\left( {{\begin{array}{*{20}c}
                 0 \hfill & 0 \hfill \\
                 \dot{x} \hfill & \dot{R} \hfill \\
              \end{array} }} \right)
 = \left( {{\begin{array}{*{20}c}
                 0 \hfill & 0 \hfill \\
                 R^T \dot{x} \hfill &  R^T \dot{R} \hfill \\
              \end{array} }} \right)
$$
Considering the pull-back of $\dot{r}, \dot{x}$ and $\varpi$ by the rotation $R$
\begin{equation}
     \dot{r}' = R^T \dot{r}, \qquad
    \dot{x}' = R^T \dot{x}, \qquad
    \varpi' = R^T \varpi
\label{pull-back of dot(x) & dot(x)_0 & varpi}
\end{equation}
and owing to (\ref{dot(R) Eulerian}), we obtain
$$ \tilde{\theta}' 
 = \left( {{\begin{array}{*{20}c}
                 0 \hfill & 0 \hfill \\
                 \dot{x}' \hfill &  j (\varpi') \hfill \\
              \end{array} }} \right)
$$
thus
$$ \vartheta_L (\dot{\mathsf{g}}) =
\left( \dot{x}', \; \; j (\varpi' ) \right)
$$
which  provides the components in the material description of the co-momentum
\begin{equation}
     \Upsilon' = \dot{x}', \qquad K' = j (\varpi' ) 
\label{Upsilon' = dot(x)' & K' = j(varpi')}
\end{equation}
$$
$$
thus, owing to (\ref{bar(omega) (a) = Upsilon + K V}), 
 (\ref{rigid body - tilde(V) = & tilde(g) = & tilde(V)' =}) and (\ref{pull-back of dot(x) & dot(x)_0 & varpi}),  we have
$$    \dot{r}'  = \bar{\theta}' (s') 
= \Upsilon' + K' \, s' 
$$
It can be verified that 
\begin{equation}
     \Pi' = p'^T, \qquad L' =  j(l'_{pr})
\label{Pi' = p'^T & L' = j (l')}
\end{equation}
$$
$$
with
\begin{equation}
   p' = R^T p, \qquad
   l'_{pr} = R^T l_{pr}
\label{pullback of p & l = l_(pr)}
\end{equation}
and the value of the dual pairing of the momentum and the co-momentum is
$$ \bm{\mu} \, \bm{\theta}  
    = p' \cdot \dot{x}' + l'_{pr} \cdot \varpi'
$$

\section{Static of arches}
\label{Section - Static of arches}

\begin{figure}[ht!]
\centering
\includegraphics[scale=1.25]{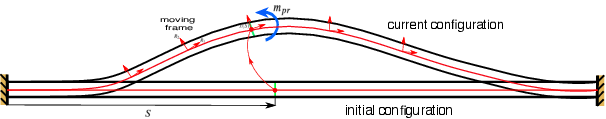}
\caption{Arch}
\label{fig Arch}
\end{figure}

Let $\mathcal{B}_3 \subset \mathbb{R}^3$ a slender body, that will be called an arch or rod, coating with matter a mean line defined in the \textit{current configuration} by a smooth map  $ \left[0, L \right] \rightarrow \mathbb{R}^3 : S \mapsto x (S)   $, where S is the arclength with respect to a given reference point of the mean line in the \textit{initial configuration} (Figure \ref{fig Arch}). The  cross-section at $x$ is the intersection of the body and the plane orthogonal to the mean line at $x$. It is supposed rigid.
For each point  $x$ of the mean line, we construct an orthonormal moving frame $(R_1, R_2, R_3)$, 
varying smoothly along the mean line. The $3 \times 3$ matrix
$$ R = (R_1 \; R_2 \;R_3)
$$
is a rotation. In a reference frame, the position $r \in \mathcal{B}_3$ of the material point $s'$ is given by
\begin{equation}
     r = R (S) \, s' + x (S)
\label{x = R (S) s' + x_0 (S)}
\end{equation}
Fixing $S$ and $s'^1 = 0$, we generate the cross-section of  $x (S)$ by varying $s'^2$ and $s'^3$.
As $R$ is a rotation, there is a column $\kappa' \in \mathbb{R}^3$
$$ \kappa' =\left( \begin{array} {c}
                       \phi'    \\
                       \kappa'_2  \\
                       \kappa'_3  \\
                    \end {array} \right)
$$
such that
\begin{equation}
     \frac{dR}{dS} = R \, j(\kappa') 
\label{dR / dS = R j (kappa')}
\end{equation}
It characterizes the arch by its angle of twist per unit length $\phi'$ and its bending curvatures  $\kappa'_2, \kappa'_3$.

To construct the co-momenta and the momenta of arches, we use once again the Maurer-Cartan 1-forms. The developments are similar to those of the dynamics of rigid bodies. Only the physical interpretation differs. 
Two viewpoints can be adopted:
\begin{itemize}
    \item \textbf{Spatial description.} 
    Later on, to simply the  notations, the derivative with respect to the arclength $S$ is denoted by the a circle surmounting the letter, then from (\ref{dR / dS = R j (kappa')}) that reads now $\mathring{R} = R \, j(\kappa')$ we deduce 
    \begin{equation}
         \mathring{R} =  j(\kappa) \, R
    \label{mathring(R) = j (kappa) R}
    \end{equation}
    with the pushforward
    $$ \kappa = R \, \kappa'
    $$
    Using the right Maurer-Cartan 1-form, we obtain by analogy  with (\ref{rigid body - Eulerian - Upsilon = & K =}) the components of the co-momentum 
    $$      \Upsilon = \mathring{x} - \kappa \times x, \qquad
   K = j (\kappa)
    $$
    Next we claim that the components of the momentum of the arch are
$$ \Pi = f^T, \qquad L =  j (m)
$$
where $f \in \mathbb{R}^3$ is the force and $m \in \mathbb{R}^3$ the moment of the internal efforts acting through the cross-section at $x$. By analogy with (\ref{mu omega = p cdot dot(x)_0 + (l - x_0 times p)}), the value of the dual pairing of the momentum and the co-momentum is
$$ \bm{\mu} \, \bm{\theta}  
    = f \cdot \mathring{x} + (m - x \times f ) \cdot \kappa
$$
where, $m$ being the moment with respect the origin of the Eulerian coordinate system, the proper moment
$$ m_{pr} = m -  x \times f 
$$
is the moment with respect to the point $x$ on the mean line defining the cross section through which the internal efforts act, then
$$ \bm{\mu} \, \bm{\theta}  
    = f \cdot \mathring{x} + m_{pr} \cdot \kappa
$$
    \item \textbf{Material description.} By pullback, we have
\begin{equation}
     \mathring{x}' = R^T \mathring{x}, \qquad
   \kappa' = R^T \kappa, \qquad
   f' = R^T f, \qquad
   m'_{pr} = R^T m_{pr}
\label{pullback of dot(x)_0 & kappa & f & m_0}
\end{equation}
Using the left Maurer-Cartan 1-form, we obtain the components of the co-momentum 
\begin{equation}
\Upsilon' = \mathring{x}', \qquad
    K' = j (\kappa')
\label{Upsilon' = ring(x)' & K' = j (kappa')}
\end{equation}
On the other hand, the components of the momentum are
\begin{equation}
 \Pi' = f'^T, \qquad L' =  j (m'_{pr})
\label{Pi' = f'^T & L' = j (m'_pr}
\end{equation}
\end{itemize}

\section{Euler-Poincar\'e equation}
\label{Section - Euler-Poincare equation}

Our aim now is to deduce by the calculus of variations the equations governing the behavior of the body (a rigid body or an arch) from an action obtained by integrating a Lagrangian $\mathfrak{L}$ depending smoothly on the co-momentum $\bm{\theta}$. The momentum associated to $\bm{\theta}$ is the dual variable given by the \textbf{constitutive relation}
\begin{equation}
     \bm{\mu} = \frac{\partial \mathfrak{L}}{\partial \bm{\theta}}
\label{mu = partial L / partial theta}
\end{equation}
which itemizes into
$$ \bm{\Pi} = \frac{\partial \mathfrak{L}}{\partial \bm{\Upsilon}_{\bm{a}_0}}, \qquad
  \bm{L}_{\bm{a}_0} = \frac{\partial \mathfrak{L}}{\partial \bm{K}}
$$
It is worth noting the remarkable fact that in the target applications the Lagrangian in the material description depends only on $\theta'$  but not explicitly on $\mathsf{g}$, 
$$ \mathfrak{L} = \mathfrak{L} (\theta')
$$
In contrast, owing to (\ref{theta = Ad (g) theta'}), the Lagrangian in the spatial description depends on both the co-momentum components $\theta$  and $\mathsf{g}$
$$ \mathfrak{L} 
= \mathfrak{L} (Ad (\mathsf{g}^{-1})\,\theta)
= \mathfrak{L} (\theta, \mathsf{g})
$$
reason for which it is more convenient to work in the material description for the variational approach. In the sequel, we suppose that
$$ \mathfrak{L} = \mathfrak{L} (\bm{\theta})
$$
When the unknown field $\mathsf{g}$  belongs to a non Abelian group, the equations of variation are Euler-Poincar\'e equations \cite{Marle 2013, Poincare 1901}. Although the method applies for the deformation of the arch as well as for the dynamics of the rigid body, we present it in the case of the dynamics. We consider the set $\mathcal{T}$  of admissible paths
$$ \left[ t_0, t_1 \right] \rightarrow \mathbb{SE} (3) : 
   t \mapsto \mathsf{g} (t)
$$
with fixed values at the extremities
\begin{equation}
     \mathsf{g} (t_0) = \mathsf{g}_0, \qquad
    \mathsf{g} (t_1) = \mathsf{g}_1
\label{g (t_0) = g_0 & g (t_1) = g_1}
\end{equation}
The associated co-momentum is the 1-form
$$ \bm{\theta} = \bm{\vartheta} (\dot{\mathsf{g}})
$$
represented in the spatial description by (\ref{theta = vartheta_R (dg) = dg g^(-1)}) and thus in the material description by (\ref{theta' = Ad(g^(-1)) theta = Ad(g^(-1)) vartheta_R(dot(g)) = vartheta_L (dot(g))}). 

The \textbf{Hamilton principle} claims that \textit{the natural path is the admissible path for which the action} 
$$ \alpha [ \mathsf{g}] = \int^{t_1}_{t_0} \mathfrak{L} (\bm{\vartheta} (\dot{\mathsf{g}})) \, \mbox{d} t
$$
\textit{is stationary}.

To deduce the equation of variations, we consider a continuous family of admissible paths 
$$ \left[ t_0, t_1 \right] \times I  \rightarrow \mathbb{SE} (3) : (t, \epsilon) \mapsto \mathsf{g} (t, \epsilon) 
= \mathsf{g}_\epsilon (t)
$$
$I$ being an interval containing zero and such that $\mathsf{g}_0$ is the natural path. Now, $\dot{\mathsf{g}}$ is a shortcut for $\partial \mathsf{g} / \partial t$ and we consider the variations 
$$ \delta \mathsf{g} = \frac{\partial \mathsf{g} }{\partial \epsilon}\mid_{\epsilon = 0}, \qquad
\delta \bm{\theta} = \frac{\partial \bm{\theta} }{\partial \epsilon}\mid_{\epsilon = 0}
$$
The stationarity condition reads
\begin{equation}
     \delta \alpha [ \mathsf{g}] = \frac{d}{dt} \alpha [ \mathsf{g}] \mid_{\epsilon = 0}
= \int^{t_1}_{t_0} \bm{\mu} \, \delta (\bm{\vartheta} (\dot{\mathsf{g}})) \, \mbox{d} t = 0
\label{delta alpha (g) = int^(t_1)_(t_0) mu delta (theta (dot(g)) dt = 0}
\end{equation}
It is represented in the material description by
\begin{equation}
     \delta \alpha [ \mathsf{g}] 
= \int^{t_1}_{t_0} \mu' \, \delta (\vartheta_L (\dot{\mathsf{g}})) \, \mbox{d} t = 0
\label{delta alpha (g) = int^(t_1)_(_0) mu' delta (vartheta_L(dot(g)) dt = 0}
\end{equation}
where 
$$ \delta (\vartheta_L (\dot{\mathsf{g}})) 
= \delta (\mathsf{g}^{-1} \dot{\mathsf{g}})
= - \mathsf{g}^{-1}  \delta \mathsf{g} \, \mathsf{g}^{-1} \dot{\mathsf{g}} 
 = - \vartheta_L (\delta \mathsf{g}) \vartheta_L (\dot{\mathsf{g}})
$$ 
and thus, by skew-symmetrization,  
the exterior derivative (denoted by a serif typeface $\mathsf{d}$) of the 1-form $\vartheta_L$ is the 2-form
$$ \mathsf{d} \vartheta_L (\delta \mathsf{g}, \dot{\mathsf{g}}) 
= \delta (\vartheta_L (\dot{\mathsf{g}}))
 - \frac{\partial}{\partial t} (\vartheta_L ( \delta (\mathsf{g})))
 = - (\vartheta_L (\delta \mathsf{g}) \vartheta_L (\dot{\mathsf{g}})
      -  \vartheta_L (\dot{\mathsf{g}}) \vartheta_L (\delta \mathsf{g}))
$$
that leads to the \textbf{Maurer-Cartan equation}
\begin{equation}
     \mathsf{d} \vartheta_L (\delta \mathsf{g}, \dot{\mathsf{g}})
= - [\vartheta_L (\delta \mathsf{g}), \vartheta_L (\dot{\mathsf{g}})]
\label{Maurer-Cartan eqns}
\end{equation}
in which $[\bullet, \bullet]$ is the bracket operator of $\mathfrak{se} (3)$ identified with the bracket of left-invariant vector fields on $\mathbb{SE} (3)$ at the identity. Using this equation, we obtain
$$  \delta (\vartheta_L (\dot{\mathsf{g}}))
 = \frac{\partial}{\partial t} (\vartheta_L ( \delta (\mathsf{g}))) + [\vartheta_L (\dot{\mathsf{g}}), \vartheta_L (\delta \mathsf{g}) ]
 = \frac{\partial}{\partial t} (\vartheta_L ( \delta (\mathsf{g}))) + ad(\vartheta_L (\dot{\mathsf{g}})) \, \vartheta_L (\delta \mathsf{g}) 
$$
in which the infinitesimal adjoint representation is denoted by $ad(\bullet) \bullet$. Introducing this expression into (\ref{delta alpha (g) = int^(t_1)_(_0) mu' delta (vartheta_L(dot(g)) dt = 0}), we have
$$      \delta \alpha [ \mathsf{g}] 
= \int^{t_1}_{t_0} \mu' \, \frac{\partial}{\partial t} (\vartheta_L ( \delta (\mathsf{g}))) \, \mbox{d} t 
 + \int^{t_1}_{t_0} \mu' \, (ad(\vartheta_L (\dot{\mathsf{g}})) \, \vartheta_L (\delta \mathsf{g})) \, \mbox{d} t = 0
$$
Using the definition of the infinitesimal coadjoint representation
$$ (ad^* (Z') \mu') \, Y' = - \mu' (ad(Z') Y')
$$
integrating by part the first term of the right hand side and owing to the conditions (\ref{g (t_0) = g_0 & g (t_1) = g_1}), we obtain
$$      \delta \alpha [ \mathsf{g}] 
= - \int^{t_1}_{t_0} \left[ \dot{\mu'}   
 +  ad^*(\vartheta_L (\dot{\mathsf{g}}) \mu' \right] \, \vartheta_L (\delta \mathsf{g})) \, \mbox{d} t = 0
$$
The variations $\vartheta_L (\delta \mathsf{g})$ being arbitrary, we obtain the \textbf{Euler-Poincar\'e equation}
\begin{equation}
     \dot{\mu'} +  ad^*(\vartheta_L (\dot{\mathsf{g}}))) \mu' = 0
\label{dot(mu)' + ad^* (theta_L (dot(g)) mu' = 0}
\end{equation}
As the Euler-Poincar\'e equation has the same form 
for all $\mathsf{g}$, we can write in coordinate-free form
$$ \dot{\bm{\mu}}   
 +  ad^*(\bm{\vartheta} (\dot{\mathsf{g}})) \bm{\mu} = 0
$$
Hence we are faced to the resolution of the following problem:

$(\mathcal{P}_1) \quad$ 
\textit{Find} $\mathsf{g}, \theta', \mu'$ \textit{such that}
\begin{equation}
     \dot{\mu'} +  ad^*(\theta') \mu' = 0
\label{dot(mu)' + ad^* (theta') mu' = 0}
\end{equation}
\begin{equation}
     \mu' = \frac{\partial \mathfrak{L}}{\partial \theta'} (\theta') = M (\theta')
\label{mu' = partial L / partial theta' = M (theta')}
\end{equation}
\begin{equation}
     \dot{\mathsf{g}} = \mathsf{g}\, \theta' \quad \mbox{with the initial condition} \quad  \mathsf{g}(0) = \mathsf{g}_0
\label{dot(g) = g theta' & g(0) = g_0}
\end{equation}

For a Lie group of dimension $n$, there are $3 \, n$ scalar unknowns for $3 \, n$ scalar equations in local charts. It is worth to remark that for a given $\theta'$, the ODE (\ref{dot(g) = g theta' & g(0) = g_0}) has always a solution. On the other hand, Equations (\ref{dot(mu)' + ad^* (theta') mu' = 0}) and (\ref{mu' = partial L / partial theta' = M (theta')}) depend on $\theta'$ and $\mu'$ but not on $\mathsf{g}$. Thus an interesting algorithm to find the natural path is to eliminate $\mu'$ between (\ref{dot(mu)' + ad^* (theta') mu' = 0}) and (\ref{mu' = partial L / partial theta' = M (theta')}) and to solve the problem according to this algorithm
\begin{itemize}
    \item Step 1: find $\theta'$ such that $ \frac{d}{dt} (M (\theta')) +  ad^*(\theta') M(\theta') = 0$
    \item Step 2: calculate $\mu'$ by (\ref{mu' = partial L / partial theta' = M (theta')})
    \item Step 3: find $\mathsf{g}$ by solving (\ref{dot(g) = g theta' & g(0) = g_0})
\end{itemize}


\subsection{Equation of motion of a rigid body}

Let us consider a rigid body of mass and moment of inertia matrix respectively given by
$$ m = \int_{\mathcal{B}'_3} \mbox{d} m (s'), \qquad
\mathcal{J}'
      = \int_{\mathcal{B}'_3} (\parallel s' \parallel^2 1_{\mathbb{R}^3} - s' s'^T)\, \mbox{d} m (s') 
$$
then by K\"onig's second theorem, the Lagrangian $\mathfrak{L}$ of a free rigid body is its kinetic energy $\mathfrak{T}$
\begin{equation}
     \mathfrak{L} (\theta')
     = \mathfrak{T}  (\theta')
  = \frac{1}{2} \, m \parallel \dot{r} \parallel^2 
  = \frac{1}{2}\, m \parallel \dot{x}' \parallel^2
                    + \frac{1}{2}\, \varpi' \cdot(\mathcal{J}'\,\varpi')
\label{rigid body - Lagrangian - material rep}
\end{equation}
and thus the components of the momentum are
\begin{equation}
      p' = \mbox{grad}_{\dot{x}'} \mathfrak{T}
      = m \, \dot{x}', \qquad
   l'_{pr} = \mbox{grad}_{\varpi'} \mathfrak{T}
   = \mathcal{J}'\,\varpi'
\label{p' = m dot(x)' & l'_pr = J' varpi'}   
\end{equation}
Owing to (\ref{coadjoint representation}), (\ref{Upsilon' = dot(x)' & K' = j(varpi')}), (\ref{Pi' = p'^T & L' = j (l')}) and (\ref{pullback of p & l = l_(pr)}), the infinitesimal coadjoint representation 
$$ \bar{\mu}' = ad^* (\theta') \, \mu'
$$
of $\mathbb{SE} (3)$ itemizes as follows
$$ \bar{p}'  = \varpi' \times p', \qquad 
   \bar{l}'_{pr} = \varpi' \times l'_{pr} + \dot{x}' \times p'
$$
and, owing to  (\ref{p' = m dot(x)' & l'_pr = J' varpi'})
$$ \bar{l}'_{pr} = \varpi' \times l'_{pr} 
$$
The Euler-Poincar\'e equation (\ref{dot(mu)' + ad^* (theta_L (dot(g)) mu' = 0}) reads
$$ \dot{p}' + \varpi' \times p' = 0, \qquad
   \dot{l}'_{pr} + \varpi' \times l'_{pr} 
   = 0
$$
where we recognize the second equation as \textbf{Euler's equation of motion} of a rigid body.
Owing to (\ref{dot(R) Eulerian}), (\ref{pull-back of dot(x) & dot(x)_0 & varpi}), (\ref{pullback of p & l = l_(pr)}), we deduce
$$      \dot{p} = 0, \qquad
   \dot{l}_{pr} = 0
$$
that gives six integrals of the motion {\it \`{a} la Poinsot}. 

\subsection{Equilibrium equations of arches}

Its Lagrangian $\mathfrak{L}$ is its elastic energy potential $\mathfrak{U}$. We consider the action 
$$ \alpha \left[ \mathsf{g} \right] 
   = \int^{L}_0 \mathfrak{L} (\theta) \, \mbox{d} S
   = \int^{L}_0 \mathfrak{U} (\theta) \, \mbox{d} S
$$
where $L$ is the length of the arch in the \textit{initial configuration}. The components of the momentum are
\begin{equation}
      f' = \mbox{grad}_{\mathring{x}'} \mathfrak{U}
      , \qquad
   m'_{pr} = \mbox{grad}_{\kappa'} \mathfrak{U}
\label{f' = grad_(ring(x)') U & l'_pr = grad_(kappa')}   
\end{equation}

Both ends of the arch are fixed. Its stationarity condition with respect to $\mathsf{g}$ entails in the material description the Euler-Poincar\'e equation
$$ \frac{d \mu'}{d S} + ad^* ({\theta'}) \, \mu' = 0
$$
In absence of the density of exterior forces and moments along the arch, the equation itemizes into
$$ \mathring{f}' + \kappa' \times f' = 0, \qquad
 \mathring{m}'_{pr} + \kappa' \times m'_{pr} + \mathring{x}' \times f' = 0
$$
We recover the \textbf{corotational equilibrium equations of arches} [4.10] et [4.11] deduced in \cite{AffineMechBook} by an engineer approach of the beam theory.
Owing to (\ref{mathring(R) = j (kappa) R}) and (\ref{pullback of dot(x)_0 & kappa & f & m_0}), a straightforward calculation leads to
$$ \mathring{f} = 0, \qquad
   \mathring{m}_{pr} + \mathring{x} \times f = 0
$$
in which we recognize the \textbf{local equilibrium equations of arches} (Equations [4.2] et [4.3] of \cite{AffineMechBook}, after multiplication by $dS / ds$ where $s$ is the arclength in the \textit{initial configuration}, or Equation (3.3) in \cite{Simo 1985}).

\section{Generalization to continuous media of arbitrary dimension}
\label{Section - Generalization to continuous media of arbitrary dimension}

So far, we have considered a material body parameterized by a single parameter (the time for the dynamics of the rigid body or the arclength for the statics of the arch). Our goal now is to generalize to bodies of which the behavior is parameterized by more than one parameter, for example the time $t$ and the arclength $S$ for the modeling of  the dynamics of an arch. 

\begin{figure}[ht!]
\centering
\includegraphics[scale=.60]{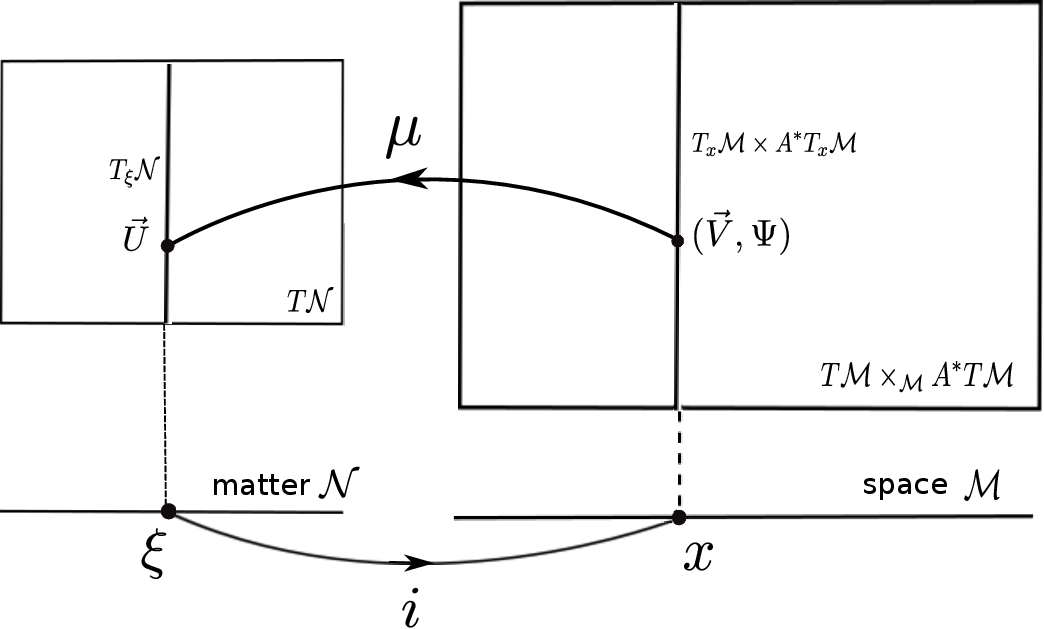}
\caption{Vector-valued Momentum}
\label{fig Vector valued momentum}
\end{figure}

\subsection{The momentum and the co-momentum in higher dimension}

For this, we must consider that the co-momentum and momentum tensors are vector-valued  (Figure \ref{fig Vector valued momentum}). The matter manifold $\mathcal{N}$ is the set of material particles $\bm{\xi}$  (with, for example, the coordinates $(t, S)$ in a local chart for the arch dynamics). We describe the motion of the matter by an embedding $i$ from $\mathcal{N}$ into the space-time $\mathcal{M}$. We consider a momentum tensor at $\bm{x} = i (\bm{\xi})$, with arguments $\vec{\bm{V}}, \bm{\Psi}$ and  value in the tangent vector space to $\mathcal{N}$ at $\bm{\xi}$. Thus we build the pullback bundle of $T\mathcal{M} \times_{\mathcal{M}} A^* T\mathcal{M}$ by moving the fiber over $\bm{x}$ to a fiber over $\bm{\xi}$, next we define the following bundle map over $\mathcal{N}$
$$ \bm{\mu}: i^* (T\mathcal{M} \times_{\mathcal{M}} A^* T\mathcal{M} ) \rightarrow T\mathcal{N} 
$$
By convention, the indices related to $\mathcal{M}$ are at the right and the material indices (related to $\mathcal{N}$) are at the left. In an affine frame $(\bm{a}_0, (\vec{\bm{e}}_\alpha))$  of $A^* T_{\bm{x}}\mathcal{M}$  and bases $(\bm{e}^\alpha)$ of $T^*_{\bm{x}} \mathcal{M}$ and $(_\gamma \vec{\bm{\eta}})$ of $T_{\bm{\xi}}\mathcal{N}$, the momentum $\bm{\mu}$ is decomposed as follows
$$ \bm{\mu} = \,_\gamma\vec{\bm{\eta}} \otimes \bm{e}^\beta \otimes\,(\,^{\gamma} \Pi_\beta \, \bm{a}_0 
              + \,^{\gamma} L^\alpha_\beta\,  \vec{\bm{e}}_\alpha)
$$
where the value of $\bm{\mu}$ is given in the basis $(_\gamma \vec{\bm{\eta}})$ and the linear combination coefficient $^\gamma \bm{\mu}$ are decomposed into the affine frame as in (\ref{decomposition of the scalar-valued momentum}). Now, $\Pi$ components have 2 indices and $L$ components have 3 indices.

Similarly, we can define a co-momentum tensor at $\bm{x} = i (\bm{\xi})$, with arguments $\bm{\Phi}, \bm{a}$,  valued in the cotangent vector space to $\mathcal{N}$ at $\bm{\xi}$ by defining the bundle map
$$ \bm{\theta}: i^* (T^*\mathcal{M} \times_{\mathcal{M}} A T\mathcal{M} ) \rightarrow T^* \mathcal{N} 
$$
In an affine frame $(\bm{a}_0, (\vec{\bm{e}}_\alpha))$  of $A^* T_{\bm{x}}\mathcal{M}$  and bases $(\bm{e}^\alpha)$ of $T^*_{\bm{x}} \mathcal{M}$ and $(_\gamma \vec{\bm{\eta}})$ of $T_{\bm{\xi}}\mathcal{N}$, the co-momentum $\bm{\theta}$ is decomposed as follows
$$ \bm{\theta} = \, ^\gamma   \bm{\eta} \otimes \vec{\mathbf{e}}_\beta \otimes (\,_\gamma \Upsilon^\beta \bm{1} + \,_\gamma K^\beta_\alpha \bm{e}^\alpha )
$$
where the value of $\bm{\theta}$ is given in the cobasis $(^\gamma \bm{\eta})$ and the linear combination coefficient $_\gamma \bm{\theta}$ are decomposed as in (\ref{decomposition of the scalar-valued co-momentum}). Now, $\Upsilon$ components have 2 indices and $K$ components have 3 indices.

Introducing the $d$-column $\,_\gamma \Upsilon$ collecting the $\,_\gamma \Upsilon^\beta$, the $d\times d$ matrix $\,_\gamma K$ of elements $\,_\gamma K^\beta_\alpha$, the $d$-row $\,^\gamma \Pi$ collecting the $\,^\gamma \Pi_\beta$ and the $d\times d$ matrix $\,^\gamma L$ of elements $\,^\gamma L^\alpha_\beta$, the dual pairing reads
$$      \bm{\mu} \, \bm{\theta} 
 = \,^\gamma \mu \,_\gamma \theta
 = (\,^\gamma \Pi, \,^\gamma L)\,  ( \,_\gamma \Upsilon, \,_\gamma K)
=  \,^\gamma \Pi \,_\gamma \Upsilon
 - \frac{1}{2} \, Tr (\,^\gamma L \,_\gamma K)
$$

By the way, the applications of this framework are not limited to arches. For the dynamics of shells, the matter manifold $\mathcal{N}$ is of dimension $3$, parameterized by the time and two curvilinear coordinates on the mean surface \cite{Boyer 2017, de Saxce 2003, de Saxce 2024}.

\subsection{Euler-Poincar\'e equation in higher dimension}

The constitutive relation (\ref{mu = partial L / partial theta}) is represented in local coordinates by
$$ \,^\gamma \Pi = \frac{\partial \mathfrak{L}}{\partial (_\gamma \Upsilon)}, \qquad
  \,^\gamma L = \frac{\partial \mathfrak{L}}{\partial (_\gamma K)}
$$
The paths are 
$$ \left[ t_0, t_1 \right] \times \left[ 0, L \right] \rightarrow \mathbb{SE} (3) : 
   (t, S) \mapsto \mathsf{g} (t, S)
$$
The action is
$$ \alpha \left[ \mathsf{g} \right] 
   = \int^{t_1}_{t_0} \int^{L}_0 \mathfrak{L} (\theta) \, \mbox{d} S \, \mbox{d} t
$$
Using the abbreviated notation 
$$ \frac{\partial}{\partial  (^\gamma  \xi)} 
= \partial_\gamma  
$$
and applying the Hamilton principle, one obtains in the material representation the generalized Euler-Poincar\'e equations
\begin{equation}
    \partial_\gamma  \,^\gamma \mu' +  ad^*(\vartheta_L (\partial_\gamma \mathsf{g}))) \,^\gamma \mu' = 0
\label{partial_gamma ^gamma mu' + ad^* (theta_L (partial_gamma g)) ^gamma  mu' = 0}
\end{equation}
When the unknown field $\mathsf{g}$ belongs to an Abelian group, the momentum tensor is divergence free
$$     \partial_\gamma  \,^\gamma \mu'  = 0
$$
When the group is not Abelian, it is not so. This issue will be discussed in Section \ref{SubSection Covariant derivative of a momentum tensor}.

Hence we are faced to the resolution of the following problem:

$(\mathcal{P}_p) \quad$ 
\textit{Find} $\mathsf{g},\; _0 \theta', \ldots, \; _{p - 1} \theta',\; ^1 \mu', \ldots, \; ^{p - 1} \mu'$ \textit{such that}
\begin{equation}
       \partial_\gamma  \,^\gamma \mu' +  ad^*(_\gamma \theta') \,^\gamma \mu' = 0
\label{partial_gamma ^gamma mu' + ad^* (_gamma theta') ^gamma mu' = 0}
\end{equation}
\begin{equation}
    ^\gamma \mu' = \frac{\partial \mathfrak{L}}{\partial _\gamma \theta'} (_0 \theta', \ldots, \; _{p - 1} \theta') =  \, ^\gamma  M (_0 \theta', \ldots, \; _{p - 1} \theta')
\label{^gamma mu' = partial L / partial _gamma theta' = ^gamma M (theta')}
\end{equation}
\begin{equation}
     \partial_\gamma \mathsf{g} = \mathsf{g} \, _\gamma \theta'
\label{partial_gamma g = g _partial gamma (gamma = 0, 1, ..., p)}
\end{equation}

For a Lie group of dimension $n$, there are $n \, (2 \, p + 1)$ scalar unknowns for $n \, (2 \, p + 1)$ scalar equations in local charts. 

We would like to adapt the 3-step method of resolution presented in Section \ref{Section - Euler-Poincare equation}. Instead of the ordinary differential equation (\ref{dot(g) = g theta' & g(0) = g_0}) , we have the system of partial differential equations (\ref{partial_gamma g = g _partial gamma (gamma = 0, 1, ..., p)}). The infinitesimal generators $\; _0 \theta', \ldots, \; _{p - 1} \theta'$ being given, in a local chart there are $n$ scalar unknowns $\mathsf{g}$ for $ p \, n $ equations, too much to be considered as independent one of each other. There is no solution unless compatibility conditions are satisfied. To obtain them, we use Frobenius's method. We must satisfy the integrability conditions
$$ \partial_\alpha (\partial_\beta \,  \mathsf{g})
 -  \partial_\beta (\partial_\alpha \mathsf{g}) = 0
$$
Owing to (\ref{partial_gamma g = g _partial gamma (gamma = 0, 1, ..., p)}) and using Leibnitz rule, one has
$$ (\partial_\alpha \mathsf{g}) _\beta \theta' 
+ \mathsf{g} \partial_\alpha (_\beta \theta' )
-  (\partial_\beta \mathsf{g}) _\alpha \theta' 
 -\mathsf{g} \partial_\beta (_\alpha \theta' ) = 0
$$
or
$$      \mathsf{g} \, (\partial_\alpha (_\beta \theta' )
  - \partial_\beta (_\alpha \theta')
  + _\alpha \theta' _\beta \theta' 
  - _\beta \theta' _\alpha \theta') = 0
$$
As the factor $\mathsf{g}$ is invertible, it can be canceled and the compatibility equations are the Maurer-Cartan equations
\begin{equation}
 \partial_\alpha (_\beta \theta' )
  - \partial_\beta (_\alpha \theta')
  = - \left[ _\alpha \theta', \,  _\beta \theta' \right]
\label{partial_alpha (_beta theta') - partial_beta (_alpha theta') = - (_alpha theta', _beta theta')}
\end{equation}

 A first algorithm is to eliminate $\mu'$ between (\ref{partial_gamma ^gamma mu' + ad^* (_gamma theta') ^gamma mu' = 0}) and (\ref{^gamma mu' = partial L / partial _gamma theta' = ^gamma M (theta')}) and to solve the problem 
\begin{itemize}
    \item Step 1: find $_0 \theta', \ldots, \; _{p - 1} \theta'$ such that 
    $$  \partial_\gamma  \, ^\gamma M +  ad^*(_\gamma \theta') \,^\gamma  M = 0
    $$
    $$ \partial_\alpha (_\beta \theta' )
  - \partial_\beta (_\alpha \theta')
  = - \left[ _\alpha \theta', \,  _\beta \theta' \right] \qquad
  (0 \leq \alpha < \beta \leq p - 1)
$$
    \item Step 2: calculate $\mu'$ by (\ref{^gamma mu' = partial L / partial _gamma theta' = ^gamma M (theta')})
    \item Step 3: find $\mathsf{g}$ by solving (\ref{partial_gamma g = g _partial gamma (gamma = 0, 1, ..., p)})
\end{itemize}

An alternative to the previous algorithm is possible if the correspondence (\ref{^gamma mu' = partial L / partial _gamma theta' = ^gamma M (theta')})  between momenta and co-momenta is one-to-one and, inspiring from Beltrami-Michell method in Elasticity, this equation can be solved with respect to the co-momentum component
\begin{equation}
    _\gamma \theta' = \;
    _\gamma T (^1 \mu', \ldots, \; ^{p - 1} \mu')
\label{_gamma theta' = = _gamma M (mu')}
\end{equation}
Then we can eliminate $\theta'$ between (\ref{partial_alpha (_beta theta') - partial_beta (_alpha theta') = - (_alpha theta', _beta theta')}) and (\ref{_gamma theta' = = _gamma M (mu')}) and to solve the alternative problem
\begin{itemize}
    \item Step 1: find $^1 \mu', \ldots, \; ^{p - 1} \mu$ such that 
    $$  \partial_\gamma  \, ^\gamma \mu' +  ad^*(_\gamma T) \,^\gamma  \mu' = 0
    $$
    $$ \partial_\alpha (_\beta T )
  - \partial_\beta (_\alpha T)
  = - \left[ _\alpha T, \,  _\beta T \right] \qquad
  (0 \leq \alpha < \beta \leq p - 1)
$$
    \item Step 2: calculate $\theta'$ by (\ref{_gamma theta' = = _gamma M (mu')})
    \item Step 3: find $\mathsf{g}$ by solving (\ref{partial_gamma g = g _partial gamma (gamma = 0, 1, ..., p)})
\end{itemize}

\subsection{Arch dynamics}

The local coordinates of the point $\bm{\xi}$ of the matter manifold are the time $\,^0 \xi = t$ and the arclength $\,^1 \xi = S $. The Lagrangian is the difference between the kinetic energy and the elastic energy potential
$$ \mathfrak{L} (\theta') = \mathfrak{T} (\dot{x}', \varpi') 
 - \mathfrak{U} (\mathring{x}', \kappa')
$$
Owing to (\ref{p' = m dot(x)' & l'_pr = J' varpi'}) and (\ref{f' = grad_(ring(x)') U & l'_pr = grad_(kappa')}), the components of the momentum are 
$$  \mbox{grad}_{\dot{x}'} \mathfrak{L} = p', \qquad
    \mbox{grad}_{\varpi'} \mathfrak{L} = l'_{pr}, \qquad
   \mbox{grad}_{\mathring{x}'} \mathfrak{L} = - f', \qquad
   \mbox{grad}_{\kappa'} \mathfrak{L} = - m'_{pr} 
$$
The generalized Euler-Poincar\'e equation (\ref{partial_gamma ^gamma mu' + ad^* (theta_L (partial_gamma g)) ^gamma  mu' = 0}) reads
$$ \mathring{f}' + \kappa' \times f' = \dot{p}' + \varpi' \times p', \qquad
   \mathring{m}'_{pr} + \kappa' \times m'_{pr} + \mathring{x}' \times f' = 
   \dot{l}'_{pr} + \varpi' \times l'_{pr} 
$$

\section{Principal connection and covariant derivative}
\label{Section - Principal connection and covariant derivative}

\begin{figure}[ht!]
\centering
\includegraphics[scale=.70]{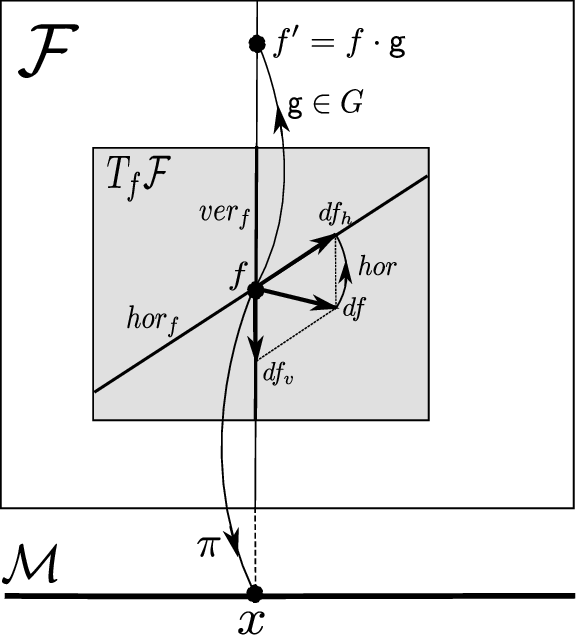}
\caption{$G$-principal bundle of affine frames}
\label{figPrincipalBundle}
\end{figure}

\subsection{Principal connection}

Let $\pi : \mathcal{F} \rightarrow \mathcal{M}$ be a $G$-principal bundle of affine frames with the free right action $(\mathsf{g}, f) \mapsto f' = f \cdot \mathsf{g}$ on each fiber (Figure \ref{figPrincipalBundle}). The corresponding infinitesimal action of $\mathfrak{g}$ is denoted $(Z, f) \mapsto df = f \cdot Z$.
Let $ver_f = Ker \,(D \pi)$ be the vertical space at $f$. An Ehresmann \textbf{principal connection} on the $G$-principle bundle $\mathcal{F}$ is a field of supplementary subspaces $hor_f$ in $T_f \mathcal{F}$ 
$$ T_f \mathcal{F} = ver_f \oplus hor_f
$$
The decomposition $df = df_v + df_h$ into a vertical vector $df_v \in ver_f $ and an horizontal vector $df_h \in hor_f $ is unique and the map $hor : T_f \mathcal{F} \rightarrow hor_f : df \mapsto df_h$ is called the horizontal projection. 

Alternatively, a connection can be defined by a field of $\mathfrak{g}$-valued $1$-forms $\Gamma$ on $\mathcal{F}$ of kernel $hor_f = Ker\, \Gamma$ such that:
\begin{itemize}
\item [$\heartsuit$] $\Gamma (f \cdot Z) = Z$, 
\item [$\spadesuit$] $\Gamma$ is $Ad$-\textbf{equivariant}: $R_\mathsf{h} \Gamma = Ad (\mathsf{h}^{-1})\,\Gamma $.
\end{itemize}
Thus $\Gamma$ is \textbf{vertical}
$$ \forall df_h\in hor_f,\qquad \Gamma (df_h) = 0
$$

Given a smooth field of vector-valued $q$-forms $\alpha$ on $\mathcal{F}$, with values in a vector space $\mathcal{V}$, for all $f \in \mathcal{F}$, the map
$$ T_f \mathcal{F} \rightarrow \mathcal{V}:
(df_1, \ldots, df_q) \mapsto 
(\alpha (f)) \, (hor (df_1), \ldots, hor (df_q))
$$
is a $q$-form which is \textbf{horizontal}, \textit{i.e.} it vanishes if one of its arguments is vertical. We denote it $\alpha_\Gamma$. The \textbf{covariant exterior derivative} of $\alpha$ is the smooth field of horizontal $(q + 1)$-forms 
$$ \mathsf{D} \alpha = (\mathsf{d} \alpha)_\Gamma
$$
The \textbf{curvature} 2-\textbf{form} of the principal connection is the covariant exterior derivative of the connection 1-form
$$ K = \mathsf{D} \Gamma
$$
It verifies the \textbf{structure equation} of the connection
\begin{equation}
     \mathsf{d} \Gamma (\delta f, \dot{f}) 
= - \left[\Gamma (\delta f), \Gamma (\dot{f})  \right ] 
+ K (\delta f, \dot{f})
\label{structure equation of a connection}
\end{equation}
If $K = 0$, we say that the connection is \textbf{flat}.

\begin{figure}[ht!]
\centering
\includegraphics[scale=.60]{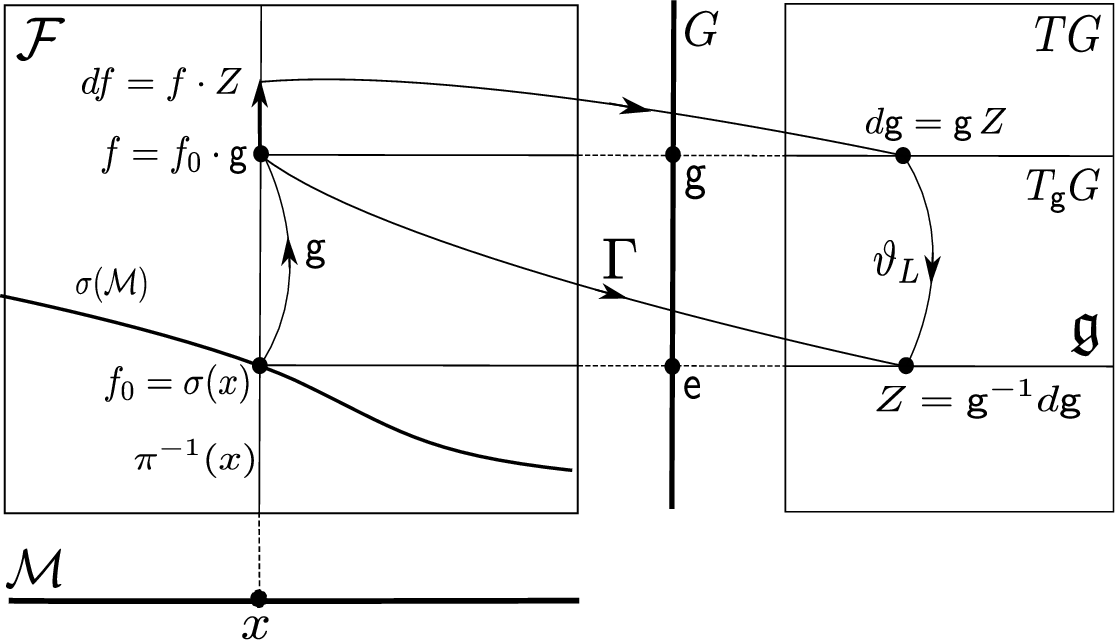}
\caption{$G$-principal connection modeled on the Maurer-Cartan 1-from}
\label{figPrincipalConnectionThetaL}
\end{figure}

Now we show how to construct a principal connection from the left Maurer-Cartan 1-form (Figure \ref{figPrincipalConnectionThetaL}). Let us pick up a smooth section $\bm{x} \mapsto f_0 = \sigma (\bm{x})$. For each frame $f$ of the  fiber over $\bm{x}$, there is a unique $\mathsf{g} \in G$ such that $f = f_0 \cdot \mathsf{g}$. By the choice of the section $\sigma$, we can identify the frame $f$ with $\mathsf{g}$ and every vertical tangent vector $df \in ver_f$ with a tangent vector $d\mathsf{g} \in T_\mathsf{g} G$. Let us show that the field of $\mathfrak{g}$-valued $1$-forms defined by
\begin{equation}
     \Gamma (df) = \vartheta_L (d\mathsf{g}) = \mathsf{g}^{-1} d\mathsf{g} 
\label{Gamma (df) = vartheta_L (dg)) = g^(-1) dg}
\end{equation}
is a principal connection. Indeed, for every $\mathsf{h} \in G , \; f \cdot \mathsf{h} = (f_0 \cdot  \mathsf{g}) \cdot \mathsf{h}  = f_0 \cdot (\mathsf{g} \, \mathsf{h})$. The tangent vector $df = f \cdot Z$ is the limit of the derivative of $f \cdot \mathsf{h}$ for $\mathsf{h} = \mathsf{e}$ in the direction $d\mathsf{h} = Z$
$$ df = f \cdot Z = f_0 \cdot (\mathsf{g} \, Z)
$$
identified with $d \mathsf{g} = \mathsf{g} \, Z$, thus we have
$$ \Gamma (f \cdot Z) = \mathsf{g}^{-1} (\mathsf{g} \, Z)
= Z
$$
that proves $\heartsuit$. Besides, we have
$$ R_\mathsf{h} \Gamma 
 = (\mathsf{g} \, \mathsf{h})^{-1} d( \mathsf{g} \, \mathsf{h})
 = \mathsf{h}^{-1} (\mathsf{g}^{-1} d\mathsf{g} ) \, \mathsf{h}
 = Ad (\mathsf{h}^{-1})\,\Gamma
$$
that proves $\spadesuit$. 

What happens if we change the section $\sigma$ of the bundle $\mathcal{F}$ by another one $\sigma'$? As the right action of $G$ on the fiber $\pi^{-1} (\bm{x})$ is transitive and free, there is a unique section $\mathsf{h}$ of $G$ such that $\sigma' (\bm{x}) = \sigma (\bm{x}) \cdot \mathsf{h} (\bm{x}) $. By the choice of the new section $\sigma'$, $f \in \pi^{-1} (\bm{x})$ is identified with $\mathsf{g}'$ and thus
$$ f = \sigma' (\bm{x}) \cdot \mathsf{g}' 
      = (\sigma (\bm{x}) \cdot \mathsf{h} (\bm{x})) \cdot \mathsf{g}' 
      = \sigma (\bm{x}) \cdot (\mathsf{h} (\bm{x})\,  \mathsf{g}' )
$$
that leads to the change of representative of $f\in \pi^{-1} (\bm{x})$ in $G$
$$ \mathsf{g} = \mathsf{h} (\bm{x})\,  \mathsf{g}' 
$$
As the 1-form $\vartheta_L$ is invariant by left translation, \textbf{the change of section has no influence on the value of the connection} (\ref{Gamma (df) = vartheta_L (dg)) = g^(-1) dg})
$$  \mathsf{g}^{-1} d\mathsf{g} 
= (\mathsf{h}\, \mathsf{g}')^{-1} d (\mathsf{h} \,\mathsf{g}') 
= \mathsf{g}'^{-1} d\mathsf{g}'
$$

Moreover, introducing in the structure equation (\ref{structure equation of a connection}) the expression of  $\delta f$ and $\dot{f}$ given by (\ref{Gamma (df) = vartheta_L (dg)) = g^(-1) dg}) and comparing to the Maurer-Cartan equation (\ref{Maurer-Cartan eqns}), we see that \textbf{the connection defined by} (\ref{Gamma (df) = vartheta_L (dg)) = g^(-1) dg}) \textbf{is flat}. Finally, for any fields of horizontal tangent vectors $\delta f, \dot{f}$ on $\mathcal{F}$,
$$  K (\delta f, \dot{f}) 
= \mathsf{d} \Gamma (\delta f, \dot{f}) 
= \delta f (\Gamma (\dot{f})) - \dot{f} (\Gamma (\delta f))
- \Gamma (\left[\delta f, \dot{f} \right]) 
$$
As the connection (\ref{Gamma (df) = vartheta_L (dg)) = g^(-1) dg}) is flat, the right hand side vanishes. As $\Gamma$ is vertical, the two former terms of the right hand side vanish too, and we have
$$ \Gamma (\left[\delta f, \dot{f} \right]) = 0
$$
thus 
$$ \delta f, \dot{f} \in hor_f \quad 
\Rightarrow \quad
\left[\delta f, \dot{f} \right] \in hor_f
$$
\textbf{For the connection} (\ref{Gamma (df) = vartheta_L (dg)) = g^(-1) dg}), \textbf{the field of horizontal vector spaces is globally integrable and defines a foliation of} $\mathcal{F}$. However, in general, the field of horizontal vector spaces is not integrable, the obstruction being the curvature. 

\subsection{Covariant derivative}
\label{SubSection - Covariant derivative}

We recall that $\pi : \mathcal{F} \rightarrow \mathcal{M}$ is a $G$-principal bundle with the free right action $(\mathsf{g}, f) \mapsto f' = f \cdot \mathsf{g}$. If $\mathcal{U}$ is a manifold on which $G$ left acts by $(\mathsf{g}, u) \mapsto u' =  \mathsf{g} \cdot u$, we define a free right action on $\mathcal{F} \times \mathcal{U}$ by
$$ (f, u) \cdot \mathsf{g} 
  = (f \cdot \mathsf{g},  \mathsf{g}^{-1} \cdot u)
$$
The orbit manifold 
$$ \mathcal{F} \times^G \mathcal{U} 
= (\mathcal{F} \times \mathcal{U})/ G
$$
exists and is called the \textbf{associated bundle} to the principal bundle $\mathcal{F}$. Its elements are the orbits
$$  \bm{u} = orb (( f, u))
$$
The \textbf{covariant derivative}  
$$ \nabla_{\overrightarrow{d \bm{x}}}\, \bm{u}
$$
of a momentum field $\bm{x} \mapsto \bm{u} (\bm{x})$ in a moving frame $\bm{x} \mapsto f (\bm{x})$ is defined by
\begin{equation}
   \nabla_{dx}\, u = du - u \cdot (\Gamma (df))
\label{tilde(nabla)_(dX) mu = d mu - (tilde(Gamma) (df)) cdot mu} 
\end{equation}
where $du$ is a shortcut for $(T u) \, dx$, that can read
\begin{equation}
   \nabla_{\overrightarrow{d \bm{x}}}\, \bm{u} = orb ((f, \nabla_{dx}\, u))
                                            = orb ((f, du - u \cdot (\Gamma (df)))
\label{tilde(nabla)_vec(dX) mu = orb (tilde(nabla)_dX rho, f) } 
\end{equation}

\section{Interpretation of the equations in terms of covariant derivatives}
\label{Section - Interpretation of the equations in terms of covariant derivatives}

\subsection{Covariant derivative of a co-momentum tensor}

In this section and the next one, we follow the framework  proposed in \cite{Castrillon 2000} and we particularize it to Euclid's group. As the component system of an Euclidean co-momentum tensor lives in the Lie algebra $\mathfrak{se} (3)$, we apply now the construction of Section \ref{SubSection - Covariant derivative} and we think of the bundle of Euclidean co-momentum tensors as modeled on the associated bundle $\mathcal{F} \times^{\mathbb{SE} (3)} \mathfrak{se} (3)$ for the free right action on $\mathcal{F} \times \mathfrak{se} (3)$ 
$$ (f, \theta) \cdot \mathsf{g} 
  = (f \cdot \mathsf{g},  Ad(\mathsf{g}^{-1}) \, \theta)
$$
where $\mathcal{F}$ is the principal bundle of Euclidean frames. By comparison with the transformation law of co-momentum tensors (\ref{theta = Ad (g) theta'}), we can identify the tensor $\bm{\theta}$ at $\bm{x}$ with the orbit
$$ \bm{\theta} = orb((f, \theta))
$$
and the bundle of co-momentum tensors with the bundle $\mathcal{F} \times^{\mathbb{SE} (3)} \mathfrak{se} (3)$.

The tangent vector $d\theta = \theta \cdot Z$ is the limit  at constant $\theta$ of the derivative
$$ d(Ad(\mathsf{h}^{-1}) \, \theta) 
= ad( d(\mathsf{h}^{-1})) \, \theta 
= - ad(\mathsf{h}^{-1} d\mathsf{h}\, \mathsf{h}^{-1}) \, \theta 
$$
for $\mathsf{h} = \mathsf{e}$ in the direction $d\mathsf{h} = Z$
$$ \theta \cdot Z = - ad(Z) \, \theta
$$
in which $df$ is identified with $d\mathsf{g} = \mathsf{g} \, Z$
$$ \theta \cdot Z = - ad(\vartheta_L (d\mathsf{g})) \, \theta
$$
Combining this result with (\ref{tilde(nabla)_(dX) mu = d mu - (tilde(Gamma) (df)) cdot mu}), the covariant derivative of the field of co-momentum tensors $\bm{\theta}$ is given by 
$$ \nabla _{dx} \, \theta = d\theta + ad(\vartheta_L (d\mathsf{g})) \, \theta
$$
In particular, for $\theta = \vartheta_L (\dot{\mathsf{g}})$ and $dx = \delta x = (T \pi) \, \delta \mathsf{g}$, one has
$$ \nabla _{\delta x} \, \vartheta_L (\dot{\mathsf{g}}) 
 =      \delta ( \vartheta_L ( \dot{\mathsf{g}}))
 +  [\vartheta_L (\delta \mathsf{g}), \vartheta_L (\dot{\mathsf{g}})] 
$$
and by swap of the variation and the time derivative 
$$ \nabla _{\dot{x}} \, \vartheta_L (\delta \mathsf{g}) 
 =      \frac{\partial}{\partial t} \vartheta_L (\delta \mathsf{g}) 
 +  [\vartheta_L (\dot{\mathsf{g}}), \vartheta_L (\delta \mathsf{g})] 
$$
The \textbf{torsion tensor} $\bm{T} $ of the connection on $\mathcal{M}$ is such that  
$$ T (\delta x, \dot{x}) 
 = \nabla _{\delta x} \, \vartheta_L (\dot{\mathsf{g}}) 
 - \nabla _{\dot{x}} \, \vartheta_L (\delta \mathsf{g})
- [\vartheta_L (\delta \mathsf{g}), \vartheta_L (\dot{\mathsf{g}})] 
$$
Owing to the skew-symmetry of the bracket and the Maurer-Cartan equation (\ref{Maurer-Cartan eqns}), we see that the  torsion vanish
$$ T (\delta x, \dot{x}) = 0
$$

\subsection{Covariant derivative of a momentum tensor}
\label{SubSection Covariant derivative of a momentum tensor}

We present now the dual version of the construction of the previous section. According to Section \ref{SubSection - Affine frames and co-frames}, we can identify an affine coframe $f^*$ with the corresponding affine frame $f$ and consider that the components of a momentum tensor are given in the frame $f$. Noting that the component system of an Euclidean momentum tensor lives in $(\mathfrak{se} (3))^*$, we define the associated bundle $\mathcal{F} \times^{\mathbb{SE} (3)} (\mathfrak{se} (3))^*$ for the free right action on $\mathcal{F} \times (\mathfrak{se} (3))^*$ 
$$ (f, \mu) \cdot \mathsf{g} 
  = (f \cdot \mathsf{g},  Ad^*(\mathsf{g}^{-1}) \, \mu)
$$
By comparison with the transformation law of Euclidean momentum tensors (\ref{mu = Ad^* (g) mu'}), we can identify the tensor $\bm{\mu}$ at $\bm{x}$ with the orbit
$$ \bm{\mu} = orb((f, \mu))
$$
and the bundle of Euclidean momentum tensors with the bundle $\mathcal{F} \times^{\mathbb{SE} (3)} (\mathfrak{se} (3))^*$.
We verify that
$$ \mu \cdot Z = - ad^*(\vartheta_L (d\mathsf{g})) \, \mu
$$
Owing to (\ref{tilde(nabla)_(dX) mu = d mu - (tilde(Gamma) (df)) cdot mu}), the covariant derivative of the field of momentum tensors $\bm{\mu}$ is given by 
$$ \nabla _{dx} \, \mu = d\mu + ad^*(\vartheta_L (d\mathsf{g})) \, \mu
$$
In particular, for the momentum $\mu'$ in the material representation, $dx = \dot{x} = (T \pi) \, \dot{\mathsf{g}}$, we have
$$ \nabla _{\dot{x}} \, \mu' = d\mu' + ad^*(\vartheta_L (\dot{\mathsf{g}})) \, \mu'
$$
that leads to the interpretation of the \textbf{Euler-Poincar\'e equation} (\ref{dot(mu)' + ad^* (theta_L (dot(g)) mu' = 0}) in terms of covariant derivatives
$$ \nabla _{\dot{x}} \, \mu' = 0
$$
In other words, \textbf{for the natural evolution, the momentum tensor is parallel-transported}. 

We generalize this equation to higher dimensions, introducing the operator of \textbf{covariant divergence} $\bm{Div} \, \bm{\mu}$  of a momentum tensor defined in an affine frame by
$$ \nabla_\gamma \,^\gamma \mu' = \partial_\gamma  \,^\gamma \mu' +  ad^*(\vartheta_L (\partial_\gamma \mathsf{g}))) \,^\gamma \mu'
$$
Hence the generalized Euler-Poincar\'e equation (\ref{partial_gamma ^gamma mu' + ad^* (theta_L (partial_gamma g)) ^gamma  mu' = 0}) means that \textbf{the field of momenta is covariant divergence free}
$$ \nabla_\gamma \,^\gamma \mu' = 0
$$

\section{Conclusions}

In this paper, we revisited the screw theory using tools of differential geometry and the tensor calculus extended to affine objects, the Euclidean co-momentum and momentum tensors of which the counterparts in screw theory are respectively the twist and the wrench. We established also the relation with the Euler-Poincar\'e equation and interpreted it in terms of parallel-transport of the momentum tensor thanks to the concept of Ehresmann connection on the principal bundle of affine frames. We showed that the left Maurer-Cartan 1-form defines a connection of which the curvature 2-form is null. In the future, we hope to extend the present work to problems of mechanics with a non-vanishing curvature as the gravitation in Einsteinian and Galilean general relativity \cite{AffineMechBook, Cardall 2024}.



\end{document}